\documentclass[aps,pra,twocolumn,showpacs,superscriptaddress]{revtex4-1}
\usepackage[latin1]{inputenc}
\usepackage{graphicx}
\usepackage{epstopdf}
\usepackage{epsfig}
\usepackage{xfrac}
\usepackage{indentfirst}
\usepackage{amsmath, amsthm, amssymb} 
\usepackage{setspace}
\usepackage{bm}
\input epsf
\usepackage{float}
\usepackage[colorlinks=true, a4paper=true, pdfstartview=FitV, linkcolor=blue, citecolor=blue, urlcolor=blue]{hyperref}
\usepackage{braket}
\begin{document}

\title{Excitation of high orbital angular momentum Rydberg states with Laguerre-Gauss beams}

\author{J. D. Rodrigues}
\affiliation{Instituto de Plasmas e Fus\~{a}o Nuclear, Instituto Superior T\'{e}cnico, Universidade de Lisboa, 1049-001 Lisbon, Portugal}
\author{L. G. Marcassa}
\affiliation{Instituto de F\'{i}sica de S\~{a}o Carlos, Universidade de S\~{a}o Paulo, Caixa Postal 369, 13560-970, S\~{a}o Carlos, S\~{a}o Paulo, Brasil}
\author{J. T. Mendon\c ca}
\affiliation{Instituto de Plasmas e Fus\~{a}o Nuclear, Instituto Superior T\'{e}cnico, Universidade de Lisboa, 1049-001 Lisbon, Portugal}


\begin{abstract}
\begin{footnotesize}

We consider the excitation of Rydberg states through photons carrying an intrinsic orbital angular momentum degree of freedom. Laguerre-Gauss modes, with a helical wave-front structure, correspond to such a set of laser beams, which carry $ \ell_0$ units of orbital angular momentum in their propagation direction, with $\ell_0$ the winding number. We demonstrate that, in a proper geometry setting, this orbital angular momentum can be transferred to the internal degrees of freedom of the atoms, thus violating the standard dipolar selection rules. Higher orbital angular momentum states become accessible through a single photon excitation process. We investigate how the spacial structure of the Laguerre-Gauss beam affects the radial coupling strength, assuming the simplest case of hydrogen-like wavefunctions. Finally we discuss a generalization of the angular momentum coupling, in order to include the effects of the fine and hyperfine splitting, in the context of the Wigner-Eckart theorem.

\end{footnotesize}
\end{abstract}

\maketitle

\section{Introduction}\label{sec:S1}

\par
Rydberg atoms \cite{gallagher, Marcassa, quantum_information1, rev1, rev2} are amongst the most exciting and promising research topics nowadays. Much of this interest arises from the particularities of these systems, such as strong dipole-dipole interactions and long lifetimes. The dipole blockade \cite{dipole1, dipole2, dipole3, dipole4} is an example of the implications of strong interactions and, applications to quantum information \cite{quantum_information1} were proposed based on possibility of performing quantum gate operations based on this effect \cite{quantum_information2, quantum_information3, quantum_information4}. These unique features become even more evident when we consider the propagation of light in Rydberg gases. Particularly, photon-photon interactions can be tailored in the blockade regime, under the conditions of electromagnetically induced transparency \cite{intro1}, leading, for instance, to the formation of Coulomb-like bound states of photons \cite{intro2}. 
\par
Until now and in most experiments, the Rydberg atoms were excited using dipole allowed transitions, either from the ground state or from an intermediate excited state. Only recently, electric dipole-forbidden transitions to Rydberg $nD$ states were observed in Rubidium atoms \cite{intro6}. Molecular resonances involving two Rydberg atoms, also called microdimers, were observed and associated with quadrupole-quadrupole interactions \cite{intro7, intro8}. A dipole-quadrupole molecular resonance was observed recently by Deiglmayr and co-workers in Caesium \cite{intro9}. All these experiments have something in common, they require a high intensity laser beam to observe such transitions since their rates are very small. It would be interesting to increase such rate in order to manipulate the properties of Rydberg atom interactions. In a recent work, Schmiegelow and co-workers have used Laguerre-Gauss laser beams to study a quadrupole transition in a Calcium ion \cite{OAM_Ca}. In principle such a technique could be applied to cold Rydberg atoms as well. 
\par
In this work we investigate the excitation of Rydberg states using orbital angular momentum (OAM) carrying photons, namely Laguerre-Gauss (LG) laser beams. As Allen and co-workers discovered in 1992, LG light fields carry a discrete amount of orbital angular momentum per photon \cite{allen1, allen2}. LG photons have been employed to encode quantum information in a higher dimensional Hilbert space \cite{quantum1}. Moreover, quantum bits (qubits) encoded in the OAM degrees of freedom of photons can lead to the violation of the no-cloning theorem \cite{intro10}. Quantum cryptography \cite{quantum2} and quantum memories \cite{quantum3} are also been considered for OAM based qubits. Various methods have been used to produce the helical wavefront structure particular of the LG beams. These techniques include spiral phase plates \cite{method1}, cylindrical lenses \cite{allen2}, forked holograms generated by computer-controlled spatial light modulators \cite{method2} or digital micro-mirror devices \cite{method3}. Good reviews of the angular momentum properties of light, as well as their practical and theoretical applications are given in Ref. \cite{review1, review2}.
\par
Concerning the interaction of LG beams with matter,the optical OAM was shown, for instance, to induce rotations and, consequently, the nucleation of vortice lattices in Bose-Einstein condensates \cite{intro3}, or entangle the photonic and atomic OAM degrees of freedom, also in cold atoms \cite{intro5}. In a plasma, for instance, this interaction was shown to promote the excitation of twisted electron plasma waves \cite{tito1}, or to enable wakefield acceleration of positrons and electrons \cite{tito2}. Recently, it was proposed that, in the case of photo-ionization of neutral atoms, the angular momentum of the photon could be transferred to the electron external degrees of freedom \cite{picon1, picon2}.
\par
In Sec. (\ref{sec:S2}), we begin by revisiting the standard dipolar selection rules for plane wave excitation. We then introduce the Laguerre-Gauss modes as an appropriate set of solutions of the Helmholtz equation in the paraxial approximation. We show, in a particular geometrical setting, that a relaxation of the standard selection rules occurs, thus increasing the number of accessible states through a single photon excitation process. The transition matrix element is decomposed into an angular and radial coupling and we show in Sec. (\ref{sec:S3}) how the later is corrected by the spacial profile of the LG laser beam. Explicit numerical calculations are presented for an hydrogen-like atom, which captures the essential ingredients of the OAM. In Section (\ref{sec:S4}), we discuss the nature of the angular momentum coupling between the LG photon and the atom in a hyperfine structure basis.  Finally, in Sec. (\ref{sec:S5}) and (\ref{sec:S6}), we discuss the results and possible applications and state some conclusions.
\section{Dipolar selection rules}\label{sec:S2}
Photon excitation processes usually rely on the interaction of the atomic dipole moment, $\boldsymbol{\mu} = -e \boldsymbol{r}$, where $\boldsymbol{r}$ is the distance of the electron to the positively charged nucleus, and the electric field of a laser beam, $\boldsymbol{\mathcal{E}} = \boldsymbol{\epsilon} \mathcal{E}_0  S\left(\boldsymbol{r}\right) e^{i\omega t}$. In this expression $\boldsymbol{\epsilon}$ is the unit polarization vector of the laser field and $S\left(\boldsymbol{r}\right)$ the spacial and phase structure of the beam, as determined by the Helmholtz equation
\begin{equation}\label{eq:helmholtz}
\left( \nabla^2 + k^2 \right) S\left( \boldsymbol{r} \right) = 0,
\end{equation}
with wavenumber $k$. 
\subsection{Plane wave excitation}
For a plane wave we simply have $S\left(\boldsymbol{r}\right) = e^{-ikr}$. In the usual dipole approximation, $e^{-ikr} \simeq 1$, we simply have $\boldsymbol{\mathcal{E}} \simeq \boldsymbol{\epsilon} \mathcal{E}_0  e^{i\omega t}$, where the size of the ground state of the atom is much smaller than the photon wavelength, $r_0 \ll \lambda$, so that the atom only senses a temporal evolution of the field. In first order perturbation theory, this time-dependent interaction, with Hamiltonian $H_{int} = \boldsymbol{\mu}\cdot \boldsymbol{\mathcal{E}}$, couples the eigenstates of the free Hamiltonian, and the relevant quantity is the dipole matrix element defined as
\begin{equation}\label{eq:matrix_element}
\mu = -e \bra{\psi_i} \boldsymbol{\epsilon} \cdot \boldsymbol{r} \ket{\psi_f} = -e \bra{nm\ell} \boldsymbol{\epsilon} \cdot \boldsymbol{r} \ket{n'\ell'm'},
\end{equation}
where the initial and final quantum states are specified by the usual quantum numbers $n$, $\ell$ and $m$. This is reasonable for hydrogen-like atoms. Later, a more accurate description of the atomic state will ge given. It becomes advantageous to represent the laser polarization on a spherical basis as $\boldsymbol{\epsilon} = \epsilon_{-1} \boldsymbol{e}_{-1} + \epsilon_{0} \boldsymbol{e}_{0} + \epsilon_{+1} \boldsymbol{e}_{+1}$ with $\boldsymbol{e}_{\mp} = 1/\sqrt{2} \left(\boldsymbol{e}_{x} \mp i \boldsymbol{e}_{y} \right)$ and $\boldsymbol{e}_{0} = \boldsymbol{e}_{z}$. Note that $\left(\boldsymbol{e}_{-1}, \boldsymbol{e}_{0}, \boldsymbol{e}_{+1} \right)$ are associated with left-circularly, linear and right-circularly polarized light. In this basis, the dipole operator can be written as $\mu_q = -e r \sqrt{4 \pi/3} Y_1^q \left(\theta, \phi\right)$ with $Y_1^q \left(\theta, \phi\right)$ the spherical harmonic function with $\ell=1$ and $m=q$, corresponding to the various photon polarization states such that $q = \left(-1,0,+1 \right)$ is associated with $\left(\sigma_{-},\pi,\sigma_{+} \right)$ transitions. The transition matrix element in Eq. (\ref{eq:matrix_element}) can then be written as
\renewcommand{\arraystretch}{2}
\begin{equation}\label{eq:matrix_element2}
\begin{array}{l l}
\mu_q  &= \bra{nm\ell} -e r \sqrt{\frac{4 \pi}{3}} Y_1^q \ket{n'm'\ell'} \\
& = \sqrt{\frac{4 \pi}{3}} \bra{\ell m}  Y_1^q \ket{\ell'm'} \bra{n\ell} -er \ket{n'\ell'}.
\end{array}
\end{equation}
The angular coupling, in the first term of this expression, is given in terms of a Clebsch-Gordan coefficients as
\begin{equation}\label{eq:CG_first}
\braket{\ell m \vert Y_L^M \vert \ell' m'} =  \sqrt{\frac{\left( 2\ell + 1 \right) \left( 2L + 1\right)}{4 \pi \left( 2 \ell' + 1 \right)}} C_{m, M, m'}^{\ell, L, \ell'} C_{0, 0, 0'}^{\ell, L, \ell'},
\end{equation}
which determines the strength of the angular momentum coupling, in this case $(\ell,m) + (1,q)  \rightarrow  (\ell',m')$. From the properties of the Clebsch-Gordan coefficients for angular momentum addition we can easily realize that for such a process as the one in Eq. (\ref{eq:matrix_element}) and (\ref{eq:matrix_element2}), the angular coupling results in $\Delta m = \pm 1, 0$ and $\Delta l = \pm 1$, the usual dipolar selection rules, with $\Delta m = m' - m$ and $\Delta \ell = \ell' - \ell$.
\begin{figure}[t!]
\includegraphics[scale=0.35]{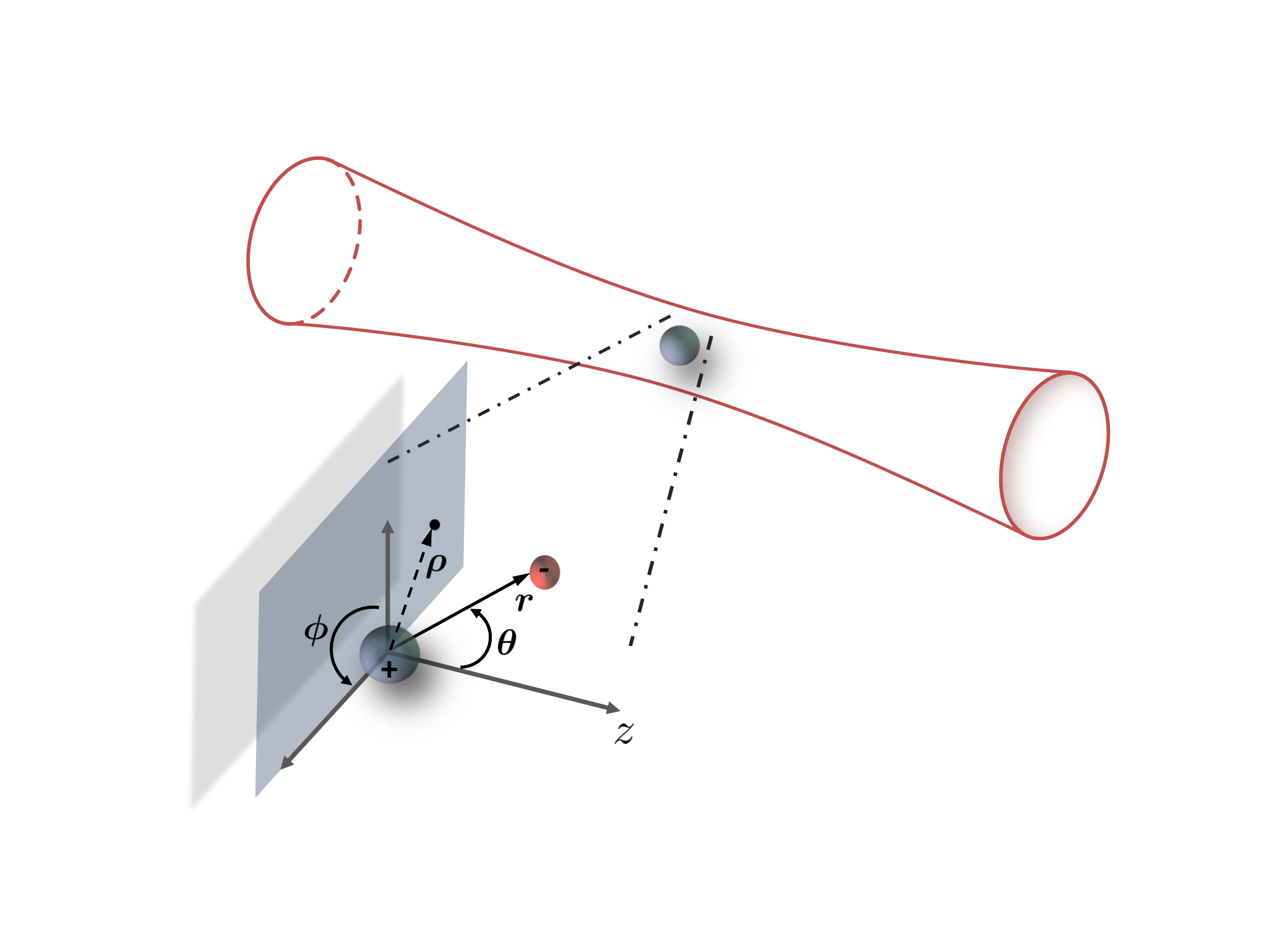}
\caption{(color online) Geometry of the excitation scheme. The atom in the ground state is placed approximately in the vortex of the Laguerre-Gauss beam, in order to take advantage of the $e^{i\ell_0 \phi}$ phase factor. We make use of a hybrid cylindrical, $\left( \rho, \phi', z \right)$, and spherical, $\left( r, \theta, \phi \right)$ coordinate system. In a coincident reference frame, the two systems are related by $\rho = r \sin \theta$, $\phi' = \phi$ and $z = r \cos \theta$ the direction of propagation of the laser beam.}
\label{fig_scheme}
\end{figure}
\subsection{Laguerre Gauss Excitation}
\begin{figure}[t!]
\includegraphics[scale=0.7]{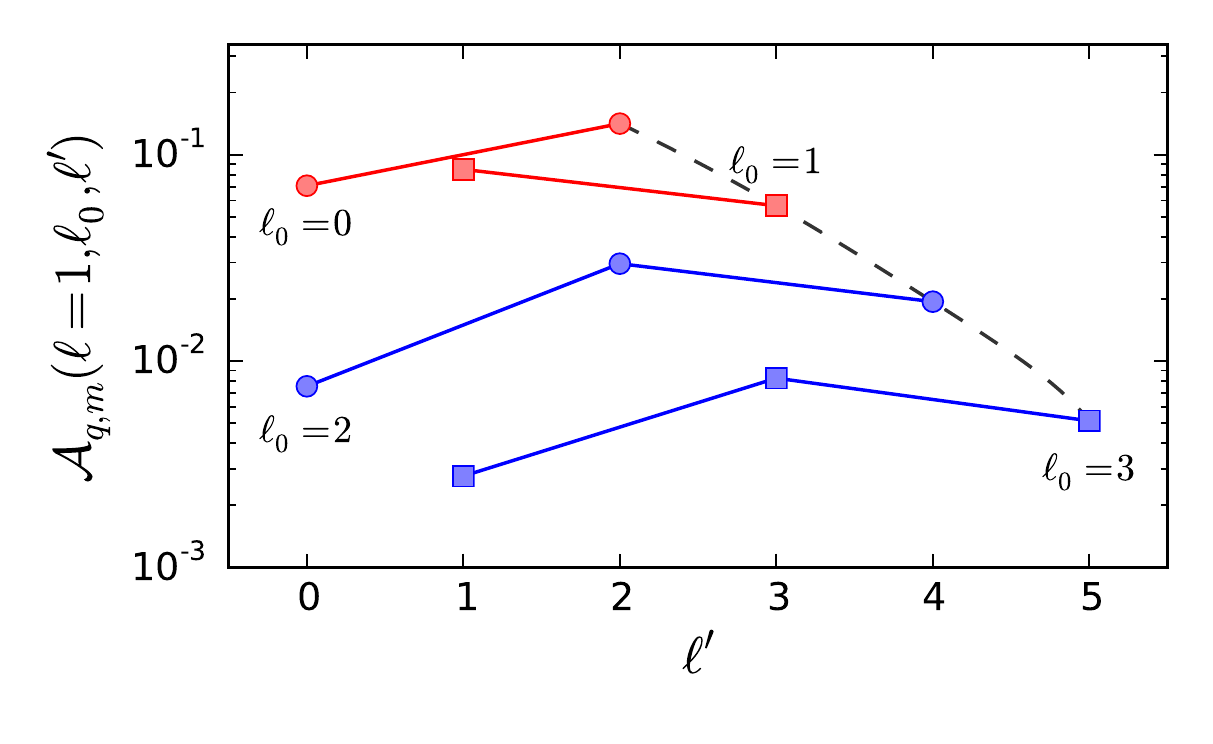}
\caption{(color online) Relative angular coupling strengths from the $P$ $\left( \ell = 1 \right)$ orbital to different final atomic states, given by $\mathcal{A}_{q,m} \left(\ell=1, \ell_0, \ell' \right)$,  where we summed over the photon polarization states, $q=-1, 0, +1$ and the initial $m=-1,0,1$ states. The different photon angular momentum states are displayed is the plot. Note that the case $\ell_0=0$ simply corresponds to the usual plane wave excitation. The final states that are not shown here correspond to forbidden dipole transitions.}
\label{fig_selection_rules}
\end{figure}

Let us now consider the case where the atomic excitation is achieved using a Laguerre-Gauss beam in a geometrical setting as the one described in Fig. (\ref{fig_scheme}). Such modes appear as solutions of the Helmholtz equation [\ref{eq:helmholtz}] in the paraxial approximation. To capture the essence of this approximation let us write the structure function as $S(\boldsymbol{r}) = s(r_{\perp}, z) e^{-ikz}$. Under the condition of strong paraxial propagation, namely $\lvert \partial^2 s / \partial z^2 \rvert \ll \lvert k \partial s / \partial z \rvert$, which corresponds to a slowly-varying amplitude function $s(r_{\perp}, z)$ along propagation direction $z$, the Helmholtz equation reduces to
\begin{equation}\label{eq:helmholtz_paraxial}
\left( \nabla_{\perp}^2 + 2ik \frac{\partial}{\partial z} \right) s\left( r_{\perp} \right) = 0.
\end{equation}
In this case, the structure function $S\left(\boldsymbol{r} \right)$, in the dipole approximation and for a given $\left(p_0, \ell_0 \right)$ mode, is determined by \cite{LG_expression, LG_expression2}
\begin{equation}\label{eq:LG}
S_{p_0 \ell_0} \left(\boldsymbol{r} \right) =  C_{p_0 \ell_0}  \zeta^{\mid \ell_0 \mid} L_{p_0}^{\mid \ell_0 \mid} \left( \zeta^2 \right) e^{-\zeta^2/2} e^{i \ell_0 \phi}
\end{equation}
with the integers $p_0$ and $\ell_0$ being the radial and azimuthal mode numbers, respectively, $\zeta \equiv \sqrt{2} \rho / w_0$ and $w_0$ the beam waist. The normalization constants $C_{p_0\ell_0}$ are given by $C_{p_0\ell_0} = \sqrt{\frac{2p_0!}{\pi \left( \vert \ell_0 \vert + p_0 \right) !}}$, and the usual Laguerre polynomials
\begin{equation}
L_{p_0}^{\ell_0} \left(x \right) = \frac{e^{x}}{p_0! x^{\ell_0}} \frac{d^{p_0}}{dx^{p_0}} \left[ x^{\ell_0+p_0} e^{-x} \right].
\end{equation}
Note that we are ignoring the Gouy phase factor, and the variation of the beam waist and the beam curvature radius along the $z$ direction, which is reasonable in the dipole approximation. Moreover, these corrections are not important to the effects we are trying to capture here. The winding number, $\ell_0$, defines the amount of OAM carried by each photon, in this case $\hbar \ell_0 $. Typically, the spread of ground state wave-function is much smaller than the beam waist and, we can approximate the dipole matrix element as
\begin{equation}\label{eq:matrix_element_LG1}
\mu = -e \bra{\psi_i} \boldsymbol{\epsilon} \cdot \boldsymbol{r} \sqrt{\frac{2}{\pi \vert \ell_0 \vert !}} \left( \frac{\sqrt{2}r\sin \theta}{w_0}\right)^{\vert \ell_0 \vert} e^{i\ell_0 \phi} \ket{\psi_f}
\end{equation}
with $r\sin \theta = \rho$. In the same way as before, $\boldsymbol{\epsilon} \cdot \boldsymbol{r} = r\sqrt{4 \pi / 3}Y_1^q$. The angular momentum properties of the LG beam are encoded in the term $\sin \theta ^{\vert \ell_0 \vert} e^{i \ell_0 \phi}$ which can be written in terms of spherical harmonics as  $\sin \theta ^{\vert \ell_0 \vert} e^{i \ell_0 \phi} = \left(2\sqrt{\frac{2\pi}{3}}\right)^{\vert \ell_0 \vert} \left[ Y_1^{\text{sgn}(\ell_0)}\right]^{\vert \ell_0 \vert}$ with $\text{sgn}(x)$ the sign function. Let us now rewrite the matrix element in Eq. (\ref{eq:matrix_element_LG1}) as
\begin{equation}\label{eq:step0}
\begin{array}{l l}
\mu = -e \sqrt{\frac{8}{3 \vert \ell_0 \vert !}} & \bra{\ell m} Y_1^q \left[ 2\sqrt{\frac{2\pi}{3}} Y_1^{\text{sgn}(\ell_0)}\right]^{\vert \ell_0 \vert} \ket{\ell'm'} \times \\
 &  \bra{n\ell} r \left(\frac{\sqrt{2}r}{w_0} \right)^{\vert \ell_0 \vert} \ket{n'\ell'},
 \end{array}
\end{equation}
where we separated the matrix element into an angular and radial coupling. We shall now explore the nature of the object $Y_1^q \left[Y_1^{\text{sgn}(\ell_0)}\right]^{\lvert \ell_0 \rvert}$. Let us begin with the OAM term $\left[Y_1^{\text{sgn}(\ell_0)}\right]^{\lvert \ell_0 \rvert}$, which corresponds to the $\lvert \ell_0 \rvert$ power of the simple spherical harmonic $Y_1^{\text{sgn}(\ell_0)}$. Using the known result for the multiplication of spherical harmonics (see Appendix \ref{sec:A2}) we obtain
\begin{equation}\label{eq:step1}
\left[ 2\sqrt{\frac{2\pi}{3}} Y_1^{\text{sgn}(\ell_0)}\right]^{\vert \ell_0 \vert} = 2^{\rvert \ell_0 \rvert} \rvert \ell_0 \rvert ! \sqrt{\frac{4 \pi}{\left( 2 \lvert \ell_0 \rvert + 1 \right) !}}  Y_{\lvert \ell_0 \rvert}^{\ell_0}
\end{equation}
which, apart from a normalization factor, simply corresponds to an angular momentum state with $\ell = \rvert \ell_0 \rvert$ and $m = \ell_0$, as expected. Let us now focus on the joint effect of the photon polarization and orbital angular momentum by investigating the full operator $Y_1^q Y_{\lvert \ell_0 \rvert}^{\ell_0}$. Ignoring for now the normalization constants, for the sake of simplicity, and proceeding in the same way as before (see Appendix \ref{sec:A2}) we can write
\begin{equation}\label{eq:step2}
Y_1^q Y_{\lvert \ell_0 \rvert}^{\ell_0} = \displaystyle\sum_{\ell^*} m \left( q, \ell_0, \ell^* \right) Y_{\ell^*}^{q + \ell_0},
\end{equation}
with
\begin{equation}\label{eq:step3}
m \left( q, \ell_0, \ell^* \right) = \sqrt{\frac{3 \left( 2 \lvert \ell_0 \rvert + 1 \right)}{4 \pi \left( 2 \ell^* + 1 \right)}} C_{q, \ell_0, q+\ell_0}^{1,\lvert \ell_0 \rvert, \ell^*} C_{0, 0, 0}^{1,\lvert \ell_0 \rvert, \ell^*}
\end{equation}
where the sum is performed over $\text{max} \lbrace \lvert \lvert \ell_0 \rvert - 1 \rvert, \lvert \ell_0 + q \rvert \rbrace  \leq \ell^* \leq \lvert \ell_0 \rvert + 1$, so that the Clebsch-Gordan coefficient in Eq. (\ref{eq:step3}) is non-vanishing. Putting together Eq. (\ref{eq:step0}), (\ref{eq:step1}), (\ref{eq:step2}) and (\ref{eq:step3}) allows us to write
\begin{equation}\label{eq:matrix_element_final_1}
\begin{array}{l l }
\mu = & -e \mathcal{D} \left( \ell_0 \right) \mathcal{R} \left(n, n', \ell_0, \ell, \ell' \right) \times \\   & \displaystyle\sum_{{\ell^{*}}} m \left( q, \ell_0, \ell^* \right) \braket{\ell m \vert Y_{{\ell^{*}}}^{q+\ell_0} \vert \ell' m'}
\end{array}
\end{equation}
with $\mathcal{D} \left( \ell_0 \right) = 2^{\lvert \ell_0 \rvert} \sqrt{\frac{32 \pi \lvert \ell_0 \rvert !}{3 \left( 2 \lvert \ell_0 \rvert + 1\right)!}}$ and 
\begin{equation}\label{eq:radial_coupling}
\mathcal{R} \left(n, n', \ell_0, \ell, \ell' \right) = \bra{n\ell} r \left(\frac{\sqrt{2}r}{w_0} \right)^{\vert \ell_0 \vert} \ket{n'\ell'}
\end{equation}
the radial coupling, modified by the geometry of the Laguerre-Gauss beam. Combining Eq. (\ref{eq:CG_first}) and (\ref{eq:matrix_element_final_1}) we can finally write
\begin{equation}\label{eq:matrix_element_FINAL}
\begin{array}{l l }
\mu = & -e \mathcal{D} \left( \ell_0 \right) \mathcal{R} \left(n, n', \ell_0, \ell, \ell' \right) \times \\   & \displaystyle\sum_{{\ell^{*}}} \mathcal{M}_{\ell}^{\ell'} \left( q, \ell_0, \ell^* \right) C_{m, q+\ell_0,m'}^{\ell, \ell^{*}, \ell'}
\end{array}
\end{equation}
with
\begin{equation}\label{eq:matrix_element_M}
\mathcal{M}_{\ell}^{\ell'} \left( q, \ell_0, \ell^* \right) =  m \left( q, \ell_0, \ell^* \right) \sqrt{\frac{\left(2 \ell + 1 \right) \left( 2 \ell^* + 1 \right)}{4 \pi \left( 2 \ell' + 1 \right)}} C_{0,0,0}^{\ell, \ell^{*}, \ell'}.
\end{equation}
It is the angular coupling given by the Clebsch-Gordan coefficients in Eq. (\ref{eq:matrix_element_FINAL}) that determines the selection rules of the excitation process described here. It directly follows from the conservation law for the sum of angular momentum projections that, a non-zero coefficient implies
\begin{equation}\label{eq:rule1}
\Delta m = m'-m = \ell_0 + q.
\end{equation}
Another important condition arising from the symmetry properties of the Clebsch-Gordan coefficients is
\begin{equation}\label{eq:rule2}
\vert \ell - \ell^* \vert \leq \ell' \leq \ell + \ell^*.
\end{equation}
This relation is known as the \textit{triangular condition}. Note that a third imposition arises due to the parity properties of the spherical harmonics. The wave-function of the state $\ket{\ell,m}$ has parity $P\left(Y_\ell^m \right) = \left(-1 \right)^\ell$ and, the condition of an overall $\left(+1\right)$ parity in each of the matrix element integrals implies that
\begin{equation}\label{eq:rule3}
\Delta \ell + \vert \ell^* \vert \quad \text{is even}.
\end{equation}
In all these expressions the term $\ell^*$ ranges between $\text{max} \lbrace \lvert \lvert \ell_0 \rvert - 1 \rvert, \lvert \ell_0 + q \rvert \rbrace  \leq \ell^* \leq \lvert \ell_0 \rvert + 1$.  Together, Eq. (\ref{eq:rule1}), (\ref{eq:rule2}) and (\ref{eq:rule3}) define the dipolar selection rules for the excitation process described here. If it often important, given an initial atomic state with OAM quantum number $\ell$, to understand which are the open transitions in the final $\ell'$ quantum number, independent of the atomic initial and final magnetic quantum numbers, $m$ and $m'$ respectively. In this sense we can define the quantity
\begin{equation}\label{eq:graph}
\begin{array}{l l}
\mathcal{A}_{q,m} \left(\ell, \ell_0, \ell' \right) = & \displaystyle\sum_{q,m} \bigg| \mathcal{D} \left( \ell_0 \right) \displaystyle\sum_{\ell^*} \mathcal{M}_{\ell}^{\ell'} \left(q, \ell_0, \ell^* \right) \times \\
 & C_{m, q+\ell_0, m+q+\ell_0}^{\ell, \ell^*, \ell'} \bigg|^2,
\end{array}
\end{equation}
where we used the selection rule in Eq. (\ref{eq:rule1}) and eliminated the dependence on initial and final magnetic quantum numbers, as well as in the photon polarization state $q$. In Fig. (\ref{fig_selection_rules}) we plot such a function, for the case of an initial $P$ orbital. In the case of $^{85}Rb$, for instance, the Rydberg excitation process usually involves a two photon process. In the first step we excite the ground state, $5S_{1/2}$, to the excited $5P_{3/2}$ state, corresponding to the D2 line at approximately 780 nm. A second pulse can then be used to the excitation of high-lying Rydberg states.
\section{Modified radial coupling}\label{sec:S3}

\begin{figure}[t!]
\includegraphics[scale=0.7]{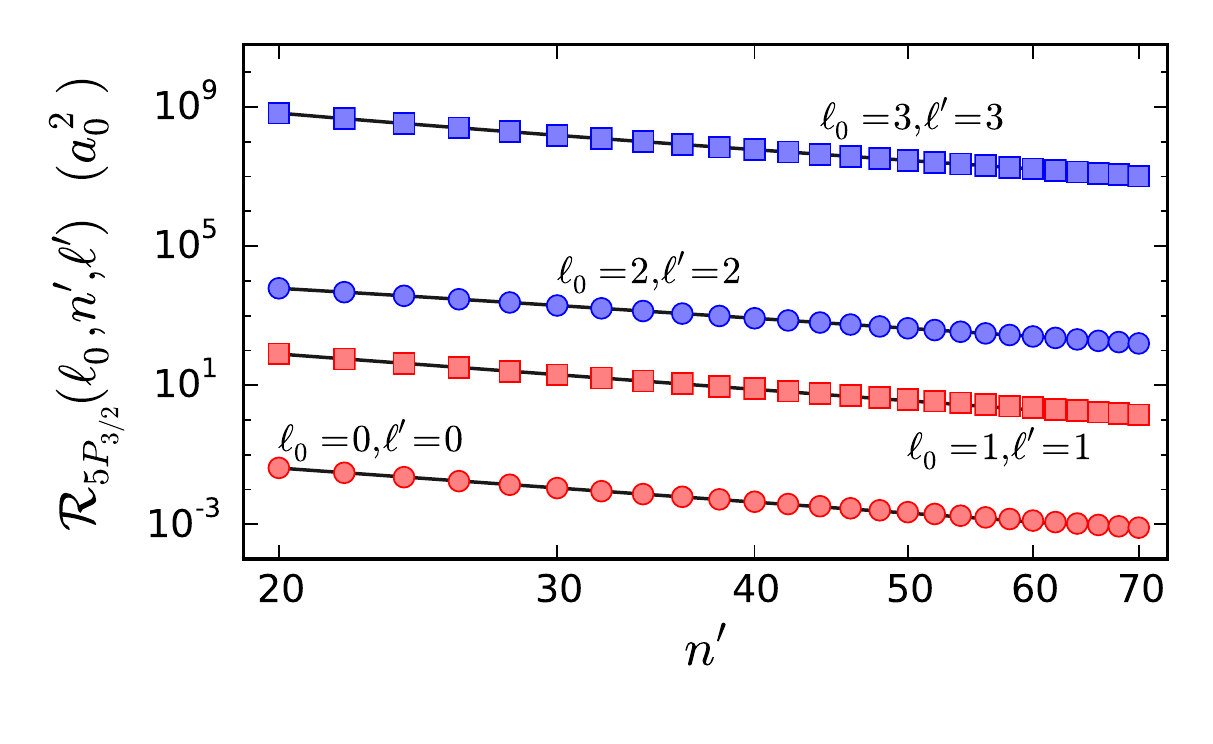}
\caption{(color online). Example of different radial couplings, as given by $\mathcal{R}_{5P_{3/2}} (\ell_0, n',\ell') = \lvert (w_0 / \sqrt{2}a_0)^{|\ell_0|}  \bra{5P_{3/2}} r^{|\ell_0| + 1}  \ket{n' \ell'}  \rvert^2$, from the initial state $\ket{5P_{3/2}}$, to different final states $\ket{n' \ell'}$ and rescaled to the ratio of the beam waist to the Bohr radius, $w_0 / (\sqrt{2} a_0)$. The different photon polarizations and final orbital angular momentum states are displayed in the plot. Each set is fitted to a scaling of the type $ C_{\ell_0, \ell_f} n ^{p}$ (black line).}
\label{fig_radial_coupling}
\end{figure}

In this section we will numerically investigate the radial coupling, $\mathcal{R}$, as defined in Rq. (\ref{eq:radial_coupling}) and, in particular, how it is modified by the transverse structure of the Laguerre-Gauss modes. For this purpose we will make use of the hydrogen atom wavefunctions, which is a good approximation for the alkali metals. In this case, analytical solutions for the normalized radial wavefunctions exist, and are given by \cite{schiff}
\begin{equation}\label{eq:wavefunction}
R_{nl} \left( r \right) = - \sqrt{\left( \frac{2}{n a_0}\right)^3 \frac{\left(n-\ell-1 \right)!}{2n \left[ \left(n+\ell \right) ! \right]^3}} e^{-r^* / 2} {r^*}^\ell L_{n+\ell}^{2\ell + 1} \left(r^* \right),
\end{equation}
with $a_0$ the Bohr radius and $r^* = \frac{2r}{n a_0}$. Later we shall discuss the validity of this approximation. In Fig. (\ref{fig_radial_coupling}) we compute the radial coupling for different sets of photon and final atomic angular momentum states. The dependence on the principal quantum number $n$ is plotted and the scaling is investigated by fitting the numerical results to a function of the type $f(n) = C_{\ell_0, \ell_f} n ^{p}$. We obtain $p=-3.2$, $-3.3$, $-2.8$ and $-2.4$, for $\ell_0 = 0$, $\ell_0 = 1$, $\ell_0 = 2$ and $\ell_0 = 3$, respectively. Note that, in the usual case of a plane wave excitation, $\ell_0 = 0$, the phenomenological scaling evidenced in experimental measures corresponds to $p=-3$ \cite{scaling}. Clearly, for higher photon OAM, the rescaled matrix element in Fig. (\ref{fig_radial_coupling}) also increases significantly, due to the higher powers of the electron position operator, $r^{\lvert \ell_0 \rvert + 1}$. Nevertheless, to obtain the important quantity relevant for experiments, we shall multiply the rescaled matrix element by the ratio of the Bohr radius to the beam waist, $\sqrt{2} a_0 / w_0$.
\par
The hydrogen wavefunctions used in the present discussion correspond to a first approximation when dealing with alkali metals, where the valence electron is immersed in a $-e/r$ Coulomb potential as the closed electron shells screen the nuclear charge. Nevertheless, for lower angular momentum states, typically $\ell \leq 3$, the electron wavefunction penetrates these closed shells and becomes exposed to the unscreened nuclear charge. This causes a deviation from the pure Coulombic potential at shorter ranges, which can be accounted for with the introduction of the quantum defect $\delta_{n\ell j}$ and an effective principal quantum number as $n^{*} = n - \delta_{n \ell j}$ \cite{gallagher}. For high orbital angular momentum states, typically $\ell > 3$, the potential is, to a high level of accuracy, a pure Coulombic potential and the quantum defects are zero. The simplified model used here allow us to capture the essentials for the effects we are trying to capture.
\section{Generalized Angular Coupling}\label{sec:S4}

We have so far assumed the atomic state to be completely described by the angular momentum quantum numbers $\ell$ and $m$ (along with the principal quantum number $n$). This allowed to tackle the problem in a simplified manner, without considering the effects of spin-orbit coupling, responsible for the fine-structure splitting of the energy levels as $\boldsymbol{J} = \boldsymbol{L} + \boldsymbol{S}$. This total electron angular momentum $\boldsymbol{J}$ also couples with the nucleus angular momentum $\boldsymbol{I}$ as $\boldsymbol{F} = \boldsymbol{J} + \boldsymbol{I}$, resulting in the hyperfine structure of the atomic levels. In this case, an atomic state is completely described by a larger set of quantum numbers as $\ket{n\ell(s)j(i)fm_f}$, where the spin and nuclear angular momentum quantum numbers, $s$ and $i$, respectively, are constant and do not couple with the photon. The question now arises of how to integrate these effects on the analysis develop so far. The answer is given by the Wigner-Eckart theorem, as it explores the symmetry properties of the angular momentum degrees of freedom and generalizes their coupling strengths. Given the result in Eq. (\ref{eq:matrix_element_final_1}), we can now write
\begin{equation}\label{eq:general2}
\begin{array}{l l}
 & \braket{\ell(s)j(i)fm_f \vert Y_1^q Y_{\lvert \ell_0 \rvert}^{\ell_0} \vert \ell'(s)j'(i)f'm'_f} = \\ & \displaystyle\sum_{\ell^*} m\left(q,\ell_0,\ell^* \right) \braket{\ell(s)j(i)fm_f \vert Y_{\ell^*}^{q+\ell_0}  \vert \ell'(s)j'(i)f'm'_f},
\end{array}
\end{equation}
where the angular coupling now depends on a larger set of quantum numbers. Note that the selection rules derived earlier, and dictated by the symmetry properties of the Clebsch-Gordan coefficients, can alternatively be determined by the Wigner-3$j$ symbols
\renewcommand{\arraystretch}{1}
\begin{equation}\label{eq:general1}
\left(
\begin{array}{ccc}
\ell & L & \ell' \\
m & M & -m'  
\end{array}
\right) = (-1)^{L-\ell-m'} \frac{1}{\sqrt{2 \ell' +1} } C_{m,M,m'}^{\ell, L, \ell'},
\end{equation}
\renewcommand{\arraystretch}{2}
which reflect the same angular momentum conservation properties. The objects $ Y_{\ell^*}^{q+\ell_0} $ constitute a set of spherical tensor operators, satisfying the commutation relations 
\begin{equation}\label{eq:wigner1}
\left[J_z, T_k^q \right] = q T_k^q \qquad \text{and}
\end{equation}
\begin{equation}\label{eq:wigner2}
\left[J_{\pm}, T_k^q \right] = \sqrt{k\left(k+1\right) - q\left(q \pm \right)} T_k^{q\pm1}.
\end{equation}
In this way, we can make use of the Wigner-Eckart theorem to write \cite{wigner3j, brink}
\renewcommand{\arraystretch}{1}
\begin{equation}\label{eq:general3}
\begin{array}{l l}
 & \braket{\ell(s)j(i)fm_f \vert Y_{\ell^*}^{q+\ell_0}  \vert \ell'(s)j'(i)f'm'_f} = \left(-1 \right)^{m'} \times \\ \\
 &  \left(
\begin{array}{ccc}
f & \ell^* & f' \\
m_f & q+\ell_0 & -m'_f   
\end{array}
\right) \braket{\ell(s)j(i)f \Vert Y_{\ell^*}  \Vert \ell'(s)j'(i)f'},
\end{array}
\end{equation}
\renewcommand{\arraystretch}{2}
where we have factorized the dependence on the magnetic quantum number $m_f$ inside the Wigner -3$j$ symbol and introduced the reduced matrix element, independent of the photon polarization state $q+\ell_0$. In the same way as before, given the properties of the Wigner-3$j$ symbols, a non-vanishing coefficient implies
\begin{equation}\label{eq:general4}
\Delta m_f = m'_f-m_f = \ell_0 + q,
\end{equation}
which is the natural generalization of the selection rule in Eq. (\ref{eq:rule1}). We shall now notice that, since the photon does not couple with the nuclear angular momentum, $\boldsymbol{I}$, we can factorize its dependence on the reduced matrix element in Eq. (\ref{eq:general3}) into a Wigner-6$j$ symbol as \cite{edmonds}
\renewcommand{\arraystretch}{1}
\begin{equation}\label{eq:general5}
\begin{array}{l l}
 & \braket{\ell(s)j(i)f \Vert Y_{\ell^*}  \Vert \ell'(s)j'(i)f'} = \left(-1\right)^{f'+j+1+i} \times \\ \\
 & \sqrt{2\left(f'+1\right)\left(2j+1\right)} \left \{
\begin{array}{ccc}
j & j' & 1 \\
f' & f & i 
\end{array}
\right \} \braket{\ell(s)j \Vert Y_{\ell^*}  \Vert \ell'(s)j'}.
\end{array}
\end{equation}
\renewcommand{\arraystretch}{2}
In the same way, given that the spin of the electron, $\boldsymbol{S}$ does not change, we can further factorize the reduced matrix element as
 \renewcommand{\arraystretch}{1}
\begin{equation}\label{eq:general6}
\begin{array}{l l}
 & \braket{\ell(s)j \Vert Y_{\ell^*}  \Vert \ell'(s)j'} = \left(-1\right)^{j'+\ell+1+s}   \times \\ \\
 &\sqrt{2\left(j'+1\right)\left(2\ell+1\right)} \left \{
\begin{array}{ccc}
\ell & \ell' & 1 \\
j' & j & s 
\end{array}
\right \} \braket{\ell\Vert Y_{\ell^*}  \Vert \ell'},
\end{array}
\end{equation}
 \renewcommand{\arraystretch}{2}
where we have defined the reduced matrix element
 \renewcommand{\arraystretch}{1}
\begin{equation}\label{eq:general7}
\braket{\ell\Vert Y_{\ell^*}  \Vert \ell'} = \sqrt{\frac{\left(2\ell+1\right)\left(2\ell^*+1\right)\left(2\ell'+1\right)}{4 \pi}} \left(
\begin{array}{ccc}
\ell & \ell^* & \ell' \\
0 & 0 & 0   
\end{array}
\right).
\end{equation}
\renewcommand{\arraystretch}{2}
Note that, by setting $\boldsymbol{I} = 0$ and $\boldsymbol{S} = 0$ in Eq. (\ref{eq:general5}) and (\ref{eq:general6}) we recover the previous results. Moreover, note that the result in Eq. (\ref{eq:general7}) allows us to recover the selection rules for quantum number $\ell$ derived in Sec. (\ref{sec:S2}), as it corresponds to the useful information when determining the accessible quantum states. In fact, and as stated before, since the electron spin and the nucleus angular momentum do not couple to the photon, the results presented in this section only introduce a correction on the angular coupling strength of the allowed transitions.
\section{Discussion}\label{sec:S5}
We shall now discuss some of the implications and limitations of the analysis develop so far. We have shown that the OAM carried by a LG photon can be transferred to the internal degrees of freedom of the atom, as suggested by recent experimental results \cite{OAM_Ca}. We considered the case where the atomic ground state is centred within the optical vortex. In fact, as long as the optical vortex is contained within the spacial extent of the ground state wavefunction, such a transfer of OAM should be expected, broadening the range of accessible quantum states as dictated by the new selection rules in Eq.(\ref{eq:rule1}), (\ref{eq:rule2}) and (\ref{eq:rule3}). Note that, in the case of a plane wave excitation, dipole-forbidden transitions can occur at much lower rates. For instance, the dipole-forbidden  $5S\rightarrow nD$, with $n = 27 \sim 59$, is observed via a quadrupole transition, although at a rate that is lower by a factor of approximately $2000$, when compared to the dipole-allowed transition $5S \rightarrow nP$, in Rubidium atoms \cite{intro6}. On the other hand, by using a LG beam with winding number $\ell_0 = 1$ and beam waist $\omega_0 \sim 1$ $\mu$m, in the geometrical setting described earlier, both the atomic transitions $5S\rightarrow nD$ and $5P\rightarrow nF$ became allowed and, moreover, we can expect them to occur at a rate which is only 10 to 100 times lower than the transition rate for $5S\rightarrow nP$ and $5P\rightarrow nD$ with a usual plane wave excitation, assuming the same electric field peak intensity $\mathcal{E}_0$. Note that the main drawback here is the vanishing of the electric field amplitude at the optical vortex and, as such, focusing the excitation beam close to the diffraction limit shall be important. This new ability to easily excite otherwise inaccessible atomic states may shed new light on the complexity of interacting Rydberg systems. For instance, in the case of molecular resonances involving two Rydberg atoms, a LG laser excitation may be used as a way to entangle the electronic and rotational degrees of freedom of the atomic pair, as suggested in \cite{intro9}.
\section{Conclusion}\label{sec:S6}
With this work we contribute to the further understanding of the interaction of orbital angular momentum bearing laser beams with matter. We considered the excitation of Rydberg states using LG laser beams and demonstrated that, in the case an atomic ground state centred at the optical vortex, orbital angular momentum can the transferred to the electron internal degrees of freedom, leading to a relaxation of the standard dipolar selection rules and allowing for the excitation of higher orbital angular momentum atomic states, at much higher rates than the ones expected for usual dipole-forbidden transitions. The modification of the radial coupling strength, due to the spacial profile of the Laguerre-Gauss beam, was investigated. We have also clarified the nature and strength of the angular coupling by discussing the effects of the fine and hyperfine splittings, based on the well known Wigner-Eckart approach. 
\par
JR acknowledges the financial support of FCT - Funda\c{c}\~{a}o da Ci\^{e}ncia e Tecnologia through the grant number SFRH/BD/52323/2013. This work was partially supported by grant 2013/02816-8, S\~{a}o Paulo Research Foundation (FAPESP), AFOSR (FA9550-12-1-0434), and CNPq.

\appendix
\section{\label{sec:A1}}

We are interested in the object $\left[ Y_1^{\text{sgn}(\ell_0)}\right]^{\vert \ell_0 \vert}$. We start from the known result for the product of two spherical harmonics, namely
\begin{equation}\label{eq:A1}
\begin{array}{l l}
Y_{\ell_1}^{m_1}Y_{\ell_2}^{m_2} =& \displaystyle\sum_{\ell} \sqrt{\frac{\left(2 \ell_1 +1 \right) \left(2 \ell_2 + 1 \right)}{4 \pi \left(2 \ell + 1 \right)}} \times  \\
 & C_{m_1, m_2, m_1 + m_2}^{\ell_1, \ell_2, \ell} C_{0, 0, 0}^{\ell_1, \ell_2, \ell} Y_{\ell}^{m_1+m_2},
 \end{array}
\end{equation}
For the case $\ell_1 = \ell_2 = m_1 = m_2 = 1$ the only non-vanishing term in the sum is $\ell = 2$ and we simply have
\begin{equation}\label{eq:A2}
Y_1^1Y_1^1 = \sqrt{\frac{(3)(3)}{4 \pi (5)}} C_{1,1,2}^{1,1,2} C_{0,0,0}^{1,1,2}  Y_2^2
\end{equation}
with $C_{1,1,2}^{1,1,2} = 1$ and $ C_{0,0,0}^{1,1,2} = \sqrt{2/3}$. Notice that, when computing a $n$-th power of the spherical harmonic $Y_1^1$, namely $\left[ Y_1^1 \right]^n$, the terms $C_{n-1-i,1,n-i}^{n-1-i,1,n-i} $ and $C_{0,0,0}^{n-1-i,1,n-i}$ appear in the product, with $i=0,...,n-2$. In this sense, we shall notice that
\begin{equation}\label{eq:A3}
\begin{array}{l l}
 & C_{n-1-i,1,n-i}^{n-1-i,1,n-i}  = 1, \quad \text{and} \\
 & C_{0,0,0}^{l,1,l+1} = \sqrt{\frac{\ell + 1}{2 \ell + 1}}.

\end{array}
\end{equation}
All the above analysis remains valid for powers of the spherical harmonics $Y_1^{-1}$. From Eq. (\ref{eq:A1}) and the properties in Eq. (\ref{eq:A3}) we can write
\begin{equation}\label{eq:A4}
\left[ Y_1^{\text{sgn}(\ell_0)}\right]^{\vert \ell_0 \vert} = \displaystyle\prod_{i=1}^{\lvert \ell_0 \rvert - 1} \sqrt{\frac{3(i+1)}{4 \pi (2i + 3)}} Y_{\lvert \ell_0 \rvert}^{\ell_0}
\end{equation}
We can now transform this particular multiplication into an expression involving factorial terms, namely
\begin{equation}\label{eq:A5}
 \left[2\sqrt{\frac{2\pi}{3}} Y_1^{\text{sgn}(\ell_0)}\right]^{\vert \ell_0 \vert} = 2^{\lvert \ell_0 \rvert} \lvert \ell_0 \rvert ! \sqrt{\frac{4 \pi}{(2 \lvert \ell_0 \rvert + 1)!}} Y_{\lvert \ell_0 \rvert}^{\ell_0} .
\end{equation}
\appendix
\section{\label{sec:A2}}

Let us compute the joint effect of photon polarization and orbital angular momentum, determined by the operator $Y_1^q Y_{\lvert \ell_0 \rvert}^{\ell_0}$. As before, we start with the expression for the product of spherical harmonics in Eq. (\ref{eq:A1}). In this case we are left with
\begin{equation}\label{eq:B1}
Y_1^q Y_{\lvert \ell_0 \rvert}^{\ell_0} = \displaystyle\sum_{\ell^*} \sqrt{\frac{3 \left( 2 \lvert \ell_0 \rvert + 1 \right)}{4 \pi \left(2 \ell^* + 1 \right)}} C_{q, \ell_0, q+\ell_0}^{1, \lvert \ell_0 \rvert, \ell^*} C_{0, 0, 0}^{1, \lvert \ell_0 \rvert, \ell^*} Y_{\ell^*}^{q+\ell_0},
\end{equation}
with $\text{max} \lbrace \lvert \lvert \ell_0 \rvert - 1 \rvert, \lvert \ell_0 + q \rvert \rbrace  \leq \ell^* \leq \lvert \ell_0 \rvert + 1$. For the sake of simplicity we write
\begin{equation}\label{eq:B2}
Y_1^q Y_{\lvert \ell_0 \rvert}^{\ell_0} = \displaystyle\sum_{\ell^*} m\left(q, \ell_0, \ell^* \right) Y_{\ell^*}^{q+\ell_0},
\end{equation}
with
\begin{equation}\label{eq:B3}
m\left(q, \ell_0, \ell^* \right) =  \sqrt{\frac{3 \left( 2 \lvert \ell_0 \rvert + 1 \right)}{4 \pi \left(2 \ell^* + 1 \right)}} C_{q, \ell_0, q+\ell_0}^{1, \lvert \ell_0 \rvert, \ell^*} C_{0, 0, 0}^{1, \lvert \ell_0 \rvert, \ell^*}.
\end{equation}

\bibliography{ref}

\begin{thebibliography}{43}%
\makeatletter
\providecommand \@ifxundefined [1]{%
 \@ifx{#1\undefined}
}%
\providecommand \@ifnum [1]{%
 \ifnum #1\expandafter \@firstoftwo
 \else \expandafter \@secondoftwo
 \fi
}%
\providecommand \@ifx [1]{%
 \ifx #1\expandafter \@firstoftwo
 \else \expandafter \@secondoftwo
 \fi
}%
\providecommand \natexlab [1]{#1}%
\providecommand \enquote  [1]{``#1''}%
\providecommand \bibnamefont  [1]{#1}%
\providecommand \bibfnamefont [1]{#1}%
\providecommand \citenamefont [1]{#1}%
\providecommand \href@noop [0]{\@secondoftwo}%
\providecommand \href [0]{\begingroup \@sanitize@url \@href}%
\providecommand \@href[1]{\@@startlink{#1}\@@href}%
\providecommand \@@href[1]{\endgroup#1\@@endlink}%
\providecommand \@sanitize@url [0]{\catcode `\\12\catcode `\$12\catcode
  `\&12\catcode `\#12\catcode `\^12\catcode `\_12\catcode `\%12\relax}%
\providecommand \@@startlink[1]{}%
\providecommand \@@endlink[0]{}%
\providecommand \url  [0]{\begingroup\@sanitize@url \@url }%
\providecommand \@url [1]{\endgroup\@href {#1}{\urlprefix }}%
\providecommand \urlprefix  [0]{URL }%
\providecommand \Eprint [0]{\href }%
\providecommand \doibase [0]{http://dx.doi.org/}%
\providecommand \selectlanguage [0]{\@gobble}%
\providecommand \bibinfo  [0]{\@secondoftwo}%
\providecommand \bibfield  [0]{\@secondoftwo}%
\providecommand \translation [1]{[#1]}%
\providecommand \BibitemOpen [0]{}%
\providecommand \bibitemStop [0]{}%
\providecommand \bibitemNoStop [0]{.\EOS\space}%
\providecommand \EOS [0]{\spacefactor3000\relax}%
\providecommand \BibitemShut  [1]{\csname bibitem#1\endcsname}%
\let\auto@bib@innerbib\@empty
\bibitem [{\citenamefont {Gallagher}(1994)}]{gallagher}%
  \BibitemOpen
  \bibfield  {author} {\bibinfo {author} {\bibfnamefont {T.~F.}\ \bibnamefont
  {Gallagher}},\ }\href@noop {} {\emph {\bibinfo {title} {Rydberg atoms}}}\
  (\bibinfo  {publisher} {Cambridge University Press},\ \bibinfo {year}
  {1994})\BibitemShut {NoStop}%
\bibitem [{\citenamefont {Marcassa}\ and\ \citenamefont
  {Shaffer}(2014)}]{Marcassa}%
  \BibitemOpen
  \bibfield  {author} {\bibinfo {author} {\bibfnamefont {L.~G.}\ \bibnamefont
  {Marcassa}}\ and\ \bibinfo {author} {\bibfnamefont {J.~P.}\ \bibnamefont
  {Shaffer}},\ }\href {\doibase
  http://dx.doi.org/10.1016/B978-0-12-800129-5.00002-X} {\ \bibinfo {series}
  {Advances In Atomic, Molecular, and Optical Physics},\ \textbf {\bibinfo
  {volume} {63}},\ \bibinfo {pages} {47 } (\bibinfo {year} {2014})}\BibitemShut
  {NoStop}%
\bibitem [{\citenamefont {Saffman}\ \emph {et~al.}(2010)\citenamefont
  {Saffman}, \citenamefont {Walker},\ and\ \citenamefont
  {M\o{}lmer}}]{quantum_information1}%
  \BibitemOpen
  \bibfield  {author} {\bibinfo {author} {\bibfnamefont {M.}~\bibnamefont
  {Saffman}}, \bibinfo {author} {\bibfnamefont {T.~G.}\ \bibnamefont {Walker}},
  \ and\ \bibinfo {author} {\bibfnamefont {K.}~\bibnamefont {M\o{}lmer}},\
  }\href {\doibase 10.1103/RevModPhys.82.2313} {\bibfield  {journal} {\bibinfo
  {journal} {Rev. Mod. Phys.}\ }\textbf {\bibinfo {volume} {82}},\ \bibinfo
  {pages} {2313} (\bibinfo {year} {2010})}\BibitemShut {NoStop}%
\bibitem [{\citenamefont {Löw}\ \emph {et~al.}(2012)\citenamefont {Löw},
  \citenamefont {Weimer}, \citenamefont {Nipper}, \citenamefont {Balewski},
  \citenamefont {Butscher}, \citenamefont {Büchler},\ and\ \citenamefont
  {Pfau}}]{rev1}%
  \BibitemOpen
  \bibfield  {author} {\bibinfo {author} {\bibfnamefont {R.}~\bibnamefont
  {Löw}}, \bibinfo {author} {\bibfnamefont {H.}~\bibnamefont {Weimer}},
  \bibinfo {author} {\bibfnamefont {J.}~\bibnamefont {Nipper}}, \bibinfo
  {author} {\bibfnamefont {J.~B.}\ \bibnamefont {Balewski}}, \bibinfo {author}
  {\bibfnamefont {B.}~\bibnamefont {Butscher}}, \bibinfo {author}
  {\bibfnamefont {H.~P.}\ \bibnamefont {Büchler}}, \ and\ \bibinfo {author}
  {\bibfnamefont {T.}~\bibnamefont {Pfau}},\ }\href
  {http://stacks.iop.org/0953-4075/45/i=11/a=113001} {\bibfield  {journal}
  {\bibinfo  {journal} {Journal of Physics B: Atomic, Molecular and Optical
  Physics}\ }\textbf {\bibinfo {volume} {45}},\ \bibinfo {pages} {113001}
  (\bibinfo {year} {2012})}\BibitemShut {NoStop}%
\bibitem [{\citenamefont {Cabral}\ \emph {et~al.}(2011)\citenamefont {Cabral},
  \citenamefont {Kondo}, \citenamefont {Gonçalves}, \citenamefont
  {Nascimento}, \citenamefont {Marcassa}, \citenamefont {Booth}, \citenamefont
  {Tallant}, \citenamefont {Schwettmann}, \citenamefont {Overstreet},
  \citenamefont {Sedlacek},\ and\ \citenamefont {Shaffer}}]{rev2}%
  \BibitemOpen
  \bibfield  {author} {\bibinfo {author} {\bibfnamefont {J.~S.}\ \bibnamefont
  {Cabral}}, \bibinfo {author} {\bibfnamefont {J.~M.}\ \bibnamefont {Kondo}},
  \bibinfo {author} {\bibfnamefont {L.~F.}\ \bibnamefont {Gonçalves}},
  \bibinfo {author} {\bibfnamefont {V.~A.}\ \bibnamefont {Nascimento}},
  \bibinfo {author} {\bibfnamefont {L.~G.}\ \bibnamefont {Marcassa}}, \bibinfo
  {author} {\bibfnamefont {D.}~\bibnamefont {Booth}}, \bibinfo {author}
  {\bibfnamefont {J.}~\bibnamefont {Tallant}}, \bibinfo {author} {\bibfnamefont
  {A.}~\bibnamefont {Schwettmann}}, \bibinfo {author} {\bibfnamefont {K.~R.}\
  \bibnamefont {Overstreet}}, \bibinfo {author} {\bibfnamefont
  {J.}~\bibnamefont {Sedlacek}}, \ and\ \bibinfo {author} {\bibfnamefont
  {J.~P.}\ \bibnamefont {Shaffer}},\ }\href
  {http://stacks.iop.org/0953-4075/44/i=18/a=184007} {\bibfield  {journal}
  {\bibinfo  {journal} {Journal of Physics B: Atomic, Molecular and Optical
  Physics}\ }\textbf {\bibinfo {volume} {44}},\ \bibinfo {pages} {184007}
  (\bibinfo {year} {2011})}\BibitemShut {NoStop}%
\bibitem [{\citenamefont {Comparat}\ and\ \citenamefont
  {Pillet}(2010)}]{dipole1}%
  \BibitemOpen
  \bibfield  {author} {\bibinfo {author} {\bibfnamefont {D.}~\bibnamefont
  {Comparat}}\ and\ \bibinfo {author} {\bibfnamefont {P.}~\bibnamefont
  {Pillet}},\ }\href {\doibase 10.1364/JOSAB.27.00A208} {\bibfield  {journal}
  {\bibinfo  {journal} {J. Opt. Soc. Am. B}\ }\textbf {\bibinfo {volume}
  {27}},\ \bibinfo {pages} {A208} (\bibinfo {year} {2010})}\BibitemShut
  {NoStop}%
\bibitem [{\citenamefont {Weidemuller}(2009)}]{dipole2}%
  \BibitemOpen
  \bibfield  {author} {\bibinfo {author} {\bibfnamefont {M.}~\bibnamefont
  {Weidemuller}},\ }\href {\doibase 10.1038/nphys1193} {\bibfield  {journal}
  {\bibinfo  {journal} {Nat Phys}\ }\textbf {\bibinfo {volume} {5}},\ \bibinfo
  {pages} {91} (\bibinfo {year} {2009})}\BibitemShut {NoStop}%
\bibitem [{\citenamefont {Urban}\ \emph {et~al.}(2009)\citenamefont {Urban},
  \citenamefont {Johnson}, \citenamefont {Henage}, \citenamefont {Isenhower},
  \citenamefont {Yavuz}, \citenamefont {Walker},\ and\ \citenamefont
  {Saffman}}]{dipole3}%
  \BibitemOpen
  \bibfield  {author} {\bibinfo {author} {\bibfnamefont {E.}~\bibnamefont
  {Urban}}, \bibinfo {author} {\bibfnamefont {T.~A.}\ \bibnamefont {Johnson}},
  \bibinfo {author} {\bibfnamefont {T.}~\bibnamefont {Henage}}, \bibinfo
  {author} {\bibfnamefont {L.}~\bibnamefont {Isenhower}}, \bibinfo {author}
  {\bibfnamefont {D.~D.}\ \bibnamefont {Yavuz}}, \bibinfo {author}
  {\bibfnamefont {T.~G.}\ \bibnamefont {Walker}}, \ and\ \bibinfo {author}
  {\bibfnamefont {M.}~\bibnamefont {Saffman}},\ }\href {\doibase
  10.1038/nphys1178} {\bibfield  {journal} {\bibinfo  {journal} {Nat Phys}\
  }\textbf {\bibinfo {volume} {5}},\ \bibinfo {pages} {110} (\bibinfo {year}
  {2009})}\BibitemShut {NoStop}%
\bibitem [{\citenamefont {Gallagher}\ and\ \citenamefont
  {Pillet}(2008)}]{dipole4}%
  \BibitemOpen
  \bibfield  {author} {\bibinfo {author} {\bibfnamefont {T.~F.}\ \bibnamefont
  {Gallagher}}\ and\ \bibinfo {author} {\bibfnamefont {P.}~\bibnamefont
  {Pillet}},\ }in\ \href {\doibase
  http://dx.doi.org/10.1016/S1049-250X(08)00013-X} {\emph {\bibinfo {booktitle}
  {Advances in Atomic, Molecular, and Optical Physics}}},\ \bibinfo {series}
  {Advances In Atomic, Molecular, and Optical Physics}, Vol.~\bibinfo {volume}
  {56}\ (\bibinfo  {publisher} {Academic Press},\ \bibinfo {year} {2008})\ pp.\
  \bibinfo {pages} {161 -- 218}\BibitemShut {NoStop}%
\bibitem [{\citenamefont {Jaksch}\ \emph {et~al.}(2000)\citenamefont {Jaksch},
  \citenamefont {Cirac}, \citenamefont {Zoller}, \citenamefont {Rolston},
  \citenamefont {C\^ot\'e},\ and\ \citenamefont
  {Lukin}}]{quantum_information2}%
  \BibitemOpen
  \bibfield  {author} {\bibinfo {author} {\bibfnamefont {D.}~\bibnamefont
  {Jaksch}}, \bibinfo {author} {\bibfnamefont {J.~I.}\ \bibnamefont {Cirac}},
  \bibinfo {author} {\bibfnamefont {P.}~\bibnamefont {Zoller}}, \bibinfo
  {author} {\bibfnamefont {S.~L.}\ \bibnamefont {Rolston}}, \bibinfo {author}
  {\bibfnamefont {R.}~\bibnamefont {C\^ot\'e}}, \ and\ \bibinfo {author}
  {\bibfnamefont {M.~D.}\ \bibnamefont {Lukin}},\ }\href {\doibase
  10.1103/PhysRevLett.85.2208} {\bibfield  {journal} {\bibinfo  {journal}
  {Phys. Rev. Lett.}\ }\textbf {\bibinfo {volume} {85}},\ \bibinfo {pages}
  {2208} (\bibinfo {year} {2000})}\BibitemShut {NoStop}%
\bibitem [{\citenamefont {Lukin}\ \emph {et~al.}(2001)\citenamefont {Lukin},
  \citenamefont {Fleischhauer}, \citenamefont {Cote}, \citenamefont {Duan},
  \citenamefont {Jaksch}, \citenamefont {Cirac},\ and\ \citenamefont
  {Zoller}}]{quantum_information3}%
  \BibitemOpen
  \bibfield  {author} {\bibinfo {author} {\bibfnamefont {M.~D.}\ \bibnamefont
  {Lukin}}, \bibinfo {author} {\bibfnamefont {M.}~\bibnamefont {Fleischhauer}},
  \bibinfo {author} {\bibfnamefont {R.}~\bibnamefont {Cote}}, \bibinfo {author}
  {\bibfnamefont {L.~M.}\ \bibnamefont {Duan}}, \bibinfo {author}
  {\bibfnamefont {D.}~\bibnamefont {Jaksch}}, \bibinfo {author} {\bibfnamefont
  {J.~I.}\ \bibnamefont {Cirac}}, \ and\ \bibinfo {author} {\bibfnamefont
  {P.}~\bibnamefont {Zoller}},\ }\href {\doibase 10.1103/PhysRevLett.87.037901}
  {\bibfield  {journal} {\bibinfo  {journal} {Phys. Rev. Lett.}\ }\textbf
  {\bibinfo {volume} {87}},\ \bibinfo {pages} {037901} (\bibinfo {year}
  {2001})}\BibitemShut {NoStop}%
\bibitem [{\citenamefont {Maller}\ \emph {et~al.}(2015)\citenamefont {Maller},
  \citenamefont {Lichtman}, \citenamefont {Xia}, \citenamefont {Sun},
  \citenamefont {Piotrowicz}, \citenamefont {Carr}, \citenamefont {Isenhower},\
  and\ \citenamefont {Saffman}}]{quantum_information4}%
  \BibitemOpen
  \bibfield  {author} {\bibinfo {author} {\bibfnamefont {K.~M.}\ \bibnamefont
  {Maller}}, \bibinfo {author} {\bibfnamefont {M.~T.}\ \bibnamefont
  {Lichtman}}, \bibinfo {author} {\bibfnamefont {T.}~\bibnamefont {Xia}},
  \bibinfo {author} {\bibfnamefont {Y.}~\bibnamefont {Sun}}, \bibinfo {author}
  {\bibfnamefont {M.~J.}\ \bibnamefont {Piotrowicz}}, \bibinfo {author}
  {\bibfnamefont {A.~W.}\ \bibnamefont {Carr}}, \bibinfo {author}
  {\bibfnamefont {L.}~\bibnamefont {Isenhower}}, \ and\ \bibinfo {author}
  {\bibfnamefont {M.}~\bibnamefont {Saffman}},\ }\href {\doibase
  10.1103/PhysRevA.92.022336} {\bibfield  {journal} {\bibinfo  {journal} {Phys.
  Rev. A}\ }\textbf {\bibinfo {volume} {92}},\ \bibinfo {pages} {022336}
  (\bibinfo {year} {2015})}\BibitemShut {NoStop}%
\bibitem [{\citenamefont {Gorshkov}\ \emph {et~al.}(2011)\citenamefont
  {Gorshkov}, \citenamefont {Otterbach}, \citenamefont {Fleischhauer},
  \citenamefont {Pohl},\ and\ \citenamefont {Lukin}}]{intro1}%
  \BibitemOpen
  \bibfield  {author} {\bibinfo {author} {\bibfnamefont {A.~V.}\ \bibnamefont
  {Gorshkov}}, \bibinfo {author} {\bibfnamefont {J.}~\bibnamefont {Otterbach}},
  \bibinfo {author} {\bibfnamefont {M.}~\bibnamefont {Fleischhauer}}, \bibinfo
  {author} {\bibfnamefont {T.}~\bibnamefont {Pohl}}, \ and\ \bibinfo {author}
  {\bibfnamefont {M.~D.}\ \bibnamefont {Lukin}},\ }\href {\doibase
  10.1103/PhysRevLett.107.133602} {\bibfield  {journal} {\bibinfo  {journal}
  {Phys. Rev. Lett.}\ }\textbf {\bibinfo {volume} {107}},\ \bibinfo {pages}
  {133602} (\bibinfo {year} {2011})}\BibitemShut {NoStop}%
\bibitem [{\citenamefont {Maghrebi}\ \emph {et~al.}(2015)\citenamefont
  {Maghrebi}, \citenamefont {Gullans}, \citenamefont {Bienias}, \citenamefont
  {Choi}, \citenamefont {Martin}, \citenamefont {Firstenberg}, \citenamefont
  {Lukin}, \citenamefont {B\"uchler},\ and\ \citenamefont {Gorshkov}}]{intro2}%
  \BibitemOpen
  \bibfield  {author} {\bibinfo {author} {\bibfnamefont {M.~F.}\ \bibnamefont
  {Maghrebi}}, \bibinfo {author} {\bibfnamefont {M.~J.}\ \bibnamefont
  {Gullans}}, \bibinfo {author} {\bibfnamefont {P.}~\bibnamefont {Bienias}},
  \bibinfo {author} {\bibfnamefont {S.}~\bibnamefont {Choi}}, \bibinfo {author}
  {\bibfnamefont {I.}~\bibnamefont {Martin}}, \bibinfo {author} {\bibfnamefont
  {O.}~\bibnamefont {Firstenberg}}, \bibinfo {author} {\bibfnamefont {M.~D.}\
  \bibnamefont {Lukin}}, \bibinfo {author} {\bibfnamefont {H.~P.}\ \bibnamefont
  {B\"uchler}}, \ and\ \bibinfo {author} {\bibfnamefont {A.~V.}\ \bibnamefont
  {Gorshkov}},\ }\href {\doibase 10.1103/PhysRevLett.115.123601} {\bibfield
  {journal} {\bibinfo  {journal} {Phys. Rev. Lett.}\ }\textbf {\bibinfo
  {volume} {115}},\ \bibinfo {pages} {123601} (\bibinfo {year}
  {2015})}\BibitemShut {NoStop}%
\bibitem [{\citenamefont {Tong}\ \emph {et~al.}(2009)\citenamefont {Tong},
  \citenamefont {Farooqi}, \citenamefont {van Kempen}, \citenamefont
  {Pavlovic}, \citenamefont {Stanojevic}, \citenamefont {C\^ot\'e},
  \citenamefont {Eyler},\ and\ \citenamefont {Gould}}]{intro6}%
  \BibitemOpen
  \bibfield  {author} {\bibinfo {author} {\bibfnamefont {D.}~\bibnamefont
  {Tong}}, \bibinfo {author} {\bibfnamefont {S.~M.}\ \bibnamefont {Farooqi}},
  \bibinfo {author} {\bibfnamefont {E.~G.~M.}\ \bibnamefont {van Kempen}},
  \bibinfo {author} {\bibfnamefont {Z.}~\bibnamefont {Pavlovic}}, \bibinfo
  {author} {\bibfnamefont {J.}~\bibnamefont {Stanojevic}}, \bibinfo {author}
  {\bibfnamefont {R.}~\bibnamefont {C\^ot\'e}}, \bibinfo {author}
  {\bibfnamefont {E.~E.}\ \bibnamefont {Eyler}}, \ and\ \bibinfo {author}
  {\bibfnamefont {P.~L.}\ \bibnamefont {Gould}},\ }\href {\doibase
  10.1103/PhysRevA.79.052509} {\bibfield  {journal} {\bibinfo  {journal} {Phys.
  Rev. A}\ }\textbf {\bibinfo {volume} {79}},\ \bibinfo {pages} {052509}
  (\bibinfo {year} {2009})}\BibitemShut {NoStop}%
\bibitem [{\citenamefont {Stanojevic}\ \emph {et~al.}(2008)\citenamefont
  {Stanojevic}, \citenamefont {C\^ot\'e}, \citenamefont {Tong}, \citenamefont
  {Eyler},\ and\ \citenamefont {Gould}}]{intro7}%
  \BibitemOpen
  \bibfield  {author} {\bibinfo {author} {\bibfnamefont {J.}~\bibnamefont
  {Stanojevic}}, \bibinfo {author} {\bibfnamefont {R.}~\bibnamefont
  {C\^ot\'e}}, \bibinfo {author} {\bibfnamefont {D.}~\bibnamefont {Tong}},
  \bibinfo {author} {\bibfnamefont {E.~E.}\ \bibnamefont {Eyler}}, \ and\
  \bibinfo {author} {\bibfnamefont {P.~L.}\ \bibnamefont {Gould}},\ }\href
  {\doibase 10.1103/PhysRevA.78.052709} {\bibfield  {journal} {\bibinfo
  {journal} {Phys. Rev. A}\ }\textbf {\bibinfo {volume} {78}},\ \bibinfo
  {pages} {052709} (\bibinfo {year} {2008})}\BibitemShut {NoStop}%
\bibitem [{\citenamefont {Schwettmann}\ \emph {et~al.}(2006)\citenamefont
  {Schwettmann}, \citenamefont {Crawford}, \citenamefont {Overstreet},\ and\
  \citenamefont {Shaffer}}]{intro8}%
  \BibitemOpen
  \bibfield  {author} {\bibinfo {author} {\bibfnamefont {A.}~\bibnamefont
  {Schwettmann}}, \bibinfo {author} {\bibfnamefont {J.}~\bibnamefont
  {Crawford}}, \bibinfo {author} {\bibfnamefont {K.~R.}\ \bibnamefont
  {Overstreet}}, \ and\ \bibinfo {author} {\bibfnamefont {J.~P.}\ \bibnamefont
  {Shaffer}},\ }\href {\doibase 10.1103/PhysRevA.74.020701} {\bibfield
  {journal} {\bibinfo  {journal} {Phys. Rev. A}\ }\textbf {\bibinfo {volume}
  {74}},\ \bibinfo {pages} {020701} (\bibinfo {year} {2006})}\BibitemShut
  {NoStop}%
\bibitem [{\citenamefont {Deiglmayr}\ \emph {et~al.}(2014)\citenamefont
  {Deiglmayr}, \citenamefont {Sa\ss{}mannshausen}, \citenamefont {Pillet},\
  and\ \citenamefont {Merkt}}]{intro9}%
  \BibitemOpen
  \bibfield  {author} {\bibinfo {author} {\bibfnamefont {J.}~\bibnamefont
  {Deiglmayr}}, \bibinfo {author} {\bibfnamefont {H.}~\bibnamefont
  {Sa\ss{}mannshausen}}, \bibinfo {author} {\bibfnamefont {P.}~\bibnamefont
  {Pillet}}, \ and\ \bibinfo {author} {\bibfnamefont {F.}~\bibnamefont
  {Merkt}},\ }\href {\doibase 10.1103/PhysRevLett.113.193001} {\bibfield
  {journal} {\bibinfo  {journal} {Phys. Rev. Lett.}\ }\textbf {\bibinfo
  {volume} {113}},\ \bibinfo {pages} {193001} (\bibinfo {year}
  {2014})}\BibitemShut {NoStop}%
\bibitem [{\citenamefont {Schmiegelow}\ \emph {et~al.}(2015)\citenamefont
  {Schmiegelow}, \citenamefont {Schulz}, \citenamefont {Kaufmann},
  \citenamefont {Ruster}, \citenamefont {Poschinger},\ and\ \citenamefont
  {Schmidt-Kaler}}]{OAM_Ca}%
  \BibitemOpen
  \bibfield  {author} {\bibinfo {author} {\bibfnamefont {C.~T.}\ \bibnamefont
  {Schmiegelow}}, \bibinfo {author} {\bibfnamefont {J.}~\bibnamefont {Schulz}},
  \bibinfo {author} {\bibfnamefont {H.}~\bibnamefont {Kaufmann}}, \bibinfo
  {author} {\bibfnamefont {T.}~\bibnamefont {Ruster}}, \bibinfo {author}
  {\bibfnamefont {U.~G.}\ \bibnamefont {Poschinger}}, \ and\ \bibinfo {author}
  {\bibfnamefont {F.}~\bibnamefont {Schmidt-Kaler}},\ }\href
  {http://arxiv.org/abs/1511.07206} {\bibfield  {journal} {\bibinfo  {journal}
  {arxiv:1511.07206}\ } (\bibinfo {year} {2015})}\BibitemShut {NoStop}%
\bibitem [{\citenamefont {Allen}\ \emph {et~al.}(1992)\citenamefont {Allen},
  \citenamefont {Beijersbergen}, \citenamefont {Spreeuw},\ and\ \citenamefont
  {Woerdman}}]{allen1}%
  \BibitemOpen
  \bibfield  {author} {\bibinfo {author} {\bibfnamefont {L.}~\bibnamefont
  {Allen}}, \bibinfo {author} {\bibfnamefont {M.~W.}\ \bibnamefont
  {Beijersbergen}}, \bibinfo {author} {\bibfnamefont {R.~J.~C.}\ \bibnamefont
  {Spreeuw}}, \ and\ \bibinfo {author} {\bibfnamefont {J.~P.}\ \bibnamefont
  {Woerdman}},\ }\href {\doibase 10.1103/PhysRevA.45.8185} {\bibfield
  {journal} {\bibinfo  {journal} {Phys. Rev. A}\ }\textbf {\bibinfo {volume}
  {45}},\ \bibinfo {pages} {8185} (\bibinfo {year} {1992})}\BibitemShut
  {NoStop}%
\bibitem [{\citenamefont {Beijersbergen}\ \emph {et~al.}(1993)\citenamefont
  {Beijersbergen}, \citenamefont {Allen}, \citenamefont {van~der Veen},\ and\
  \citenamefont {Woerdman}}]{allen2}%
  \BibitemOpen
  \bibfield  {author} {\bibinfo {author} {\bibfnamefont {M.}~\bibnamefont
  {Beijersbergen}}, \bibinfo {author} {\bibfnamefont {L.}~\bibnamefont
  {Allen}}, \bibinfo {author} {\bibfnamefont {H.}~\bibnamefont {van~der Veen}},
  \ and\ \bibinfo {author} {\bibfnamefont {J.}~\bibnamefont {Woerdman}},\
  }\href {\doibase http://dx.doi.org/10.1016/0030-4018(93)90535-D} {\bibfield
  {journal} {\bibinfo  {journal} {Optics Communications}\ }\textbf {\bibinfo
  {volume} {96}},\ \bibinfo {pages} {123 } (\bibinfo {year}
  {1993})}\BibitemShut {NoStop}%
\bibitem [{\citenamefont {Fickler}\ \emph {et~al.}(2014)\citenamefont
  {Fickler}, \citenamefont {Lapkiewicz}, \citenamefont {Huber}, \citenamefont
  {Lavery}, \citenamefont {Padgett},\ and\ \citenamefont
  {Zeilinger}}]{quantum1}%
  \BibitemOpen
  \bibfield  {author} {\bibinfo {author} {\bibfnamefont {R.}~\bibnamefont
  {Fickler}}, \bibinfo {author} {\bibfnamefont {R.}~\bibnamefont {Lapkiewicz}},
  \bibinfo {author} {\bibfnamefont {M.}~\bibnamefont {Huber}}, \bibinfo
  {author} {\bibfnamefont {M.~P.~J.}\ \bibnamefont {Lavery}}, \bibinfo {author}
  {\bibfnamefont {M.~J.}\ \bibnamefont {Padgett}}, \ and\ \bibinfo {author}
  {\bibfnamefont {A.}~\bibnamefont {Zeilinger}},\ }\href
  {http://dx.doi.org/10.1038/ncomms5502} {\bibfield  {journal} {\bibinfo
  {journal} {Nat Commun}\ }\textbf {\bibinfo {volume} {5}} (\bibinfo {year}
  {2014})},\ \bibinfo {note} {article}\BibitemShut {NoStop}%
\bibitem [{\citenamefont {Nagali}\ \emph {et~al.}(2009)\citenamefont {Nagali},
  \citenamefont {Sansoni}, \citenamefont {Sciarrino}, \citenamefont
  {De~Martini}, \citenamefont {Marrucci}, \citenamefont {Piccirillo},
  \citenamefont {Karimi},\ and\ \citenamefont {Santamato}}]{intro10}%
  \BibitemOpen
  \bibfield  {author} {\bibinfo {author} {\bibfnamefont {E.}~\bibnamefont
  {Nagali}}, \bibinfo {author} {\bibfnamefont {L.}~\bibnamefont {Sansoni}},
  \bibinfo {author} {\bibfnamefont {F.}~\bibnamefont {Sciarrino}}, \bibinfo
  {author} {\bibfnamefont {F.}~\bibnamefont {De~Martini}}, \bibinfo {author}
  {\bibfnamefont {L.}~\bibnamefont {Marrucci}}, \bibinfo {author}
  {\bibfnamefont {B.}~\bibnamefont {Piccirillo}}, \bibinfo {author}
  {\bibfnamefont {E.}~\bibnamefont {Karimi}}, \ and\ \bibinfo {author}
  {\bibfnamefont {E.}~\bibnamefont {Santamato}},\ }\href {\doibase
  10.1038/nphoton.2009.214} {\bibfield  {journal} {\bibinfo  {journal} {Nat
  Photon}\ }\textbf {\bibinfo {volume} {3}},\ \bibinfo {pages} {720} (\bibinfo
  {year} {2009})}\BibitemShut {NoStop}%
\bibitem [{\citenamefont {Souza}\ \emph {et~al.}(2008)\citenamefont {Souza},
  \citenamefont {Borges}, \citenamefont {Khoury}, \citenamefont {Huguenin},
  \citenamefont {Aolita},\ and\ \citenamefont {Walborn}}]{quantum2}%
  \BibitemOpen
  \bibfield  {author} {\bibinfo {author} {\bibfnamefont {C.~E.~R.}\
  \bibnamefont {Souza}}, \bibinfo {author} {\bibfnamefont {C.~V.~S.}\
  \bibnamefont {Borges}}, \bibinfo {author} {\bibfnamefont {A.~Z.}\
  \bibnamefont {Khoury}}, \bibinfo {author} {\bibfnamefont {J.~A.~O.}\
  \bibnamefont {Huguenin}}, \bibinfo {author} {\bibfnamefont {L.}~\bibnamefont
  {Aolita}}, \ and\ \bibinfo {author} {\bibfnamefont {S.~P.}\ \bibnamefont
  {Walborn}},\ }\href {\doibase 10.1103/PhysRevA.77.032345} {\bibfield
  {journal} {\bibinfo  {journal} {Phys. Rev. A}\ }\textbf {\bibinfo {volume}
  {77}},\ \bibinfo {pages} {032345} (\bibinfo {year} {2008})}\BibitemShut
  {NoStop}%
\bibitem [{\citenamefont {NicolasA.}\ \emph {et~al.}(2014)\citenamefont
  {NicolasA.}, \citenamefont {VeissierL.}, \citenamefont {GinerL.},
  \citenamefont {GiacobinoE.}, \citenamefont {MaxeinD.},\ and\ \citenamefont
  {LauratJ.}}]{quantum3}%
  \BibitemOpen
  \bibfield  {author} {\bibinfo {author} {\bibnamefont {NicolasA.}}, \bibinfo
  {author} {\bibnamefont {VeissierL.}}, \bibinfo {author} {\bibnamefont
  {GinerL.}}, \bibinfo {author} {\bibnamefont {GiacobinoE.}}, \bibinfo {author}
  {\bibnamefont {MaxeinD.}}, \ and\ \bibinfo {author} {\bibnamefont
  {LauratJ.}},\ }\href {http://dx.doi.org/10.1038/nphoton.2013.355} {\bibfield
  {journal} {\bibinfo  {journal} {Nat Photon}\ }\textbf {\bibinfo {volume}
  {8}},\ \bibinfo {pages} {234} (\bibinfo {year} {2014})},\ \bibinfo {note}
  {letter}\BibitemShut {NoStop}%
\bibitem [{\citenamefont {Beijersbergen}\ \emph {et~al.}(1994)\citenamefont
  {Beijersbergen}, \citenamefont {Coerwinkel}, \citenamefont {Kristensen},\
  and\ \citenamefont {Woerdman}}]{method1}%
  \BibitemOpen
  \bibfield  {author} {\bibinfo {author} {\bibfnamefont {M.}~\bibnamefont
  {Beijersbergen}}, \bibinfo {author} {\bibfnamefont {R.}~\bibnamefont
  {Coerwinkel}}, \bibinfo {author} {\bibfnamefont {M.}~\bibnamefont
  {Kristensen}}, \ and\ \bibinfo {author} {\bibfnamefont {J.}~\bibnamefont
  {Woerdman}},\ }\href {\doibase
  http://dx.doi.org/10.1016/0030-4018(94)90638-6} {\bibfield  {journal}
  {\bibinfo  {journal} {Optics Communications}\ }\textbf {\bibinfo {volume}
  {112}},\ \bibinfo {pages} {321 } (\bibinfo {year} {1994})}\BibitemShut
  {NoStop}%
\bibitem [{\citenamefont {Curtis}\ \emph {et~al.}(2002)\citenamefont {Curtis},
  \citenamefont {Koss},\ and\ \citenamefont {Grier}}]{method2}%
  \BibitemOpen
  \bibfield  {author} {\bibinfo {author} {\bibfnamefont {J.~E.}\ \bibnamefont
  {Curtis}}, \bibinfo {author} {\bibfnamefont {B.~A.}\ \bibnamefont {Koss}}, \
  and\ \bibinfo {author} {\bibfnamefont {D.~G.}\ \bibnamefont {Grier}},\ }\href
  {\doibase http://dx.doi.org/10.1016/S0030-4018(02)01524-9} {\bibfield
  {journal} {\bibinfo  {journal} {Optics Communications}\ }\textbf {\bibinfo
  {volume} {207}},\ \bibinfo {pages} {169 } (\bibinfo {year}
  {2002})}\BibitemShut {NoStop}%
\bibitem [{\citenamefont {Mirhosseini}\ \emph {et~al.}(2013)\citenamefont
  {Mirhosseini}, \citenamefont {{n}a Loaiza}, \citenamefont {Chen},
  \citenamefont {Rodenburg}, \citenamefont {Malik},\ and\ \citenamefont
  {Boyd}}]{method3}%
  \BibitemOpen
  \bibfield  {author} {\bibinfo {author} {\bibfnamefont {M.}~\bibnamefont
  {Mirhosseini}}, \bibinfo {author} {\bibfnamefont {O.~S.~M.}\ \bibnamefont
  {{n}a Loaiza}}, \bibinfo {author} {\bibfnamefont {C.}~\bibnamefont {Chen}},
  \bibinfo {author} {\bibfnamefont {B.}~\bibnamefont {Rodenburg}}, \bibinfo
  {author} {\bibfnamefont {M.}~\bibnamefont {Malik}}, \ and\ \bibinfo {author}
  {\bibfnamefont {R.~W.}\ \bibnamefont {Boyd}},\ }\href {\doibase
  10.1364/OE.21.030196} {\bibfield  {journal} {\bibinfo  {journal} {Opt.
  Express}\ }\textbf {\bibinfo {volume} {21}},\ \bibinfo {pages} {30196}
  (\bibinfo {year} {2013})}\BibitemShut {NoStop}%
\bibitem [{\citenamefont {Franke-Arnold}\ \emph {et~al.}(2008)\citenamefont
  {Franke-Arnold}, \citenamefont {Allen},\ and\ \citenamefont
  {Padgett}}]{review1}%
  \BibitemOpen
  \bibfield  {author} {\bibinfo {author} {\bibfnamefont {S.}~\bibnamefont
  {Franke-Arnold}}, \bibinfo {author} {\bibfnamefont {L.}~\bibnamefont
  {Allen}}, \ and\ \bibinfo {author} {\bibfnamefont {M.}~\bibnamefont
  {Padgett}},\ }\href {\doibase 10.1002/lpor.200810007} {\bibfield  {journal}
  {\bibinfo  {journal} {Laser and Photonics Reviews}\ }\textbf {\bibinfo
  {volume} {2}},\ \bibinfo {pages} {299} (\bibinfo {year} {2008})}\BibitemShut
  {NoStop}%
\bibitem [{\citenamefont {Yao}\ and\ \citenamefont {Padgett}(2011)}]{review2}%
  \BibitemOpen
  \bibfield  {author} {\bibinfo {author} {\bibfnamefont {A.~M.}\ \bibnamefont
  {Yao}}\ and\ \bibinfo {author} {\bibfnamefont {M.~J.}\ \bibnamefont
  {Padgett}},\ }\href {\doibase 10.1364/AOP.3.000161} {\bibfield  {journal}
  {\bibinfo  {journal} {Adv. Opt. Photon.}\ }\textbf {\bibinfo {volume} {3}},\
  \bibinfo {pages} {161} (\bibinfo {year} {2011})}\BibitemShut {NoStop}%
\bibitem [{\citenamefont {Andersen}\ \emph {et~al.}(2006)\citenamefont
  {Andersen}, \citenamefont {Ryu}, \citenamefont {Clad\'e}, \citenamefont
  {Natarajan}, \citenamefont {Vaziri}, \citenamefont {Helmerson},\ and\
  \citenamefont {Phillips}}]{intro3}%
  \BibitemOpen
  \bibfield  {author} {\bibinfo {author} {\bibfnamefont {M.~F.}\ \bibnamefont
  {Andersen}}, \bibinfo {author} {\bibfnamefont {C.}~\bibnamefont {Ryu}},
  \bibinfo {author} {\bibfnamefont {P.}~\bibnamefont {Clad\'e}}, \bibinfo
  {author} {\bibfnamefont {V.}~\bibnamefont {Natarajan}}, \bibinfo {author}
  {\bibfnamefont {A.}~\bibnamefont {Vaziri}}, \bibinfo {author} {\bibfnamefont
  {K.}~\bibnamefont {Helmerson}}, \ and\ \bibinfo {author} {\bibfnamefont
  {W.~D.}\ \bibnamefont {Phillips}},\ }\href {\doibase
  10.1103/PhysRevLett.97.170406} {\bibfield  {journal} {\bibinfo  {journal}
  {Phys. Rev. Lett.}\ }\textbf {\bibinfo {volume} {97}},\ \bibinfo {pages}
  {170406} (\bibinfo {year} {2006})}\BibitemShut {NoStop}%
\bibitem [{\citenamefont {Inoue}\ \emph {et~al.}(2006)\citenamefont {Inoue},
  \citenamefont {Kanai}, \citenamefont {Yonehara}, \citenamefont {Miyamoto},
  \citenamefont {Koashi},\ and\ \citenamefont {Kozuma}}]{intro5}%
  \BibitemOpen
  \bibfield  {author} {\bibinfo {author} {\bibfnamefont {R.}~\bibnamefont
  {Inoue}}, \bibinfo {author} {\bibfnamefont {N.}~\bibnamefont {Kanai}},
  \bibinfo {author} {\bibfnamefont {T.}~\bibnamefont {Yonehara}}, \bibinfo
  {author} {\bibfnamefont {Y.}~\bibnamefont {Miyamoto}}, \bibinfo {author}
  {\bibfnamefont {M.}~\bibnamefont {Koashi}}, \ and\ \bibinfo {author}
  {\bibfnamefont {M.}~\bibnamefont {Kozuma}},\ }\href {\doibase
  10.1103/PhysRevA.74.053809} {\bibfield  {journal} {\bibinfo  {journal} {Phys.
  Rev. A}\ }\textbf {\bibinfo {volume} {74}},\ \bibinfo {pages} {053809}
  (\bibinfo {year} {2006})}\BibitemShut {NoStop}%
\bibitem [{\citenamefont {Mendon\ifmmode~\mbox{\c{c}}\else \c{c}\fi{}a}\ \emph
  {et~al.}(2009)\citenamefont {Mendon\ifmmode~\mbox{\c{c}}\else \c{c}\fi{}a},
  \citenamefont {Thid\'e},\ and\ \citenamefont {Then}}]{tito1}%
  \BibitemOpen
  \bibfield  {author} {\bibinfo {author} {\bibfnamefont {J.~T.}\ \bibnamefont
  {Mendon\ifmmode~\mbox{\c{c}}\else \c{c}\fi{}a}}, \bibinfo {author}
  {\bibfnamefont {B.}~\bibnamefont {Thid\'e}}, \ and\ \bibinfo {author}
  {\bibfnamefont {H.}~\bibnamefont {Then}},\ }\href {\doibase
  10.1103/PhysRevLett.102.185005} {\bibfield  {journal} {\bibinfo  {journal}
  {Phys. Rev. Lett.}\ }\textbf {\bibinfo {volume} {102}},\ \bibinfo {pages}
  {185005} (\bibinfo {year} {2009})}\BibitemShut {NoStop}%
\bibitem [{\citenamefont {Vieira}\ and\ \citenamefont
  {Mendon\ifmmode~\mbox{\c{c}}\else \c{c}\fi{}a}(2014)}]{tito2}%
  \BibitemOpen
  \bibfield  {author} {\bibinfo {author} {\bibfnamefont {J.}~\bibnamefont
  {Vieira}}\ and\ \bibinfo {author} {\bibfnamefont {J.~T.}\ \bibnamefont
  {Mendon\ifmmode~\mbox{\c{c}}\else \c{c}\fi{}a}},\ }\href {\doibase
  10.1103/PhysRevLett.112.215001} {\bibfield  {journal} {\bibinfo  {journal}
  {Phys. Rev. Lett.}\ }\textbf {\bibinfo {volume} {112}},\ \bibinfo {pages}
  {215001} (\bibinfo {year} {2014})}\BibitemShut {NoStop}%
\bibitem [{\citenamefont {Picon}\ \emph
  {et~al.}(2010{\natexlab{a}})\citenamefont {Picon}, \citenamefont {and. J
  Mompart and. J. R. Vazquez~de Aldana}, \citenamefont {Plaja}, \citenamefont
  {Calvo},\ and\ \citenamefont {Roso}}]{picon1}%
  \BibitemOpen
  \bibfield  {author} {\bibinfo {author} {\bibfnamefont {A.}~\bibnamefont
  {Picon}}, \bibinfo {author} {\bibfnamefont {A.~B.}\ \bibnamefont {and. J
  Mompart and. J. R. Vazquez~de Aldana}}, \bibinfo {author} {\bibfnamefont
  {L.}~\bibnamefont {Plaja}}, \bibinfo {author} {\bibfnamefont {G.~F.}\
  \bibnamefont {Calvo}}, \ and\ \bibinfo {author} {\bibfnamefont
  {L.}~\bibnamefont {Roso}},\ }\href
  {http://stacks.iop.org/1367-2630/12/i=8/a=083053} {\bibfield  {journal}
  {\bibinfo  {journal} {New Journal of Physics}\ }\textbf {\bibinfo {volume}
  {12}},\ \bibinfo {pages} {083053} (\bibinfo {year}
  {2010}{\natexlab{a}})}\BibitemShut {NoStop}%
\bibitem [{\citenamefont {Picon}\ \emph
  {et~al.}(2010{\natexlab{b}})\citenamefont {Picon}, \citenamefont {Mompart},
  \citenamefont {de~Aldana}, \citenamefont {Plaja}, \citenamefont {Calvo},\
  and\ \citenamefont {Roso}}]{picon2}%
  \BibitemOpen
  \bibfield  {author} {\bibinfo {author} {\bibfnamefont {A.}~\bibnamefont
  {Picon}}, \bibinfo {author} {\bibfnamefont {J.}~\bibnamefont {Mompart}},
  \bibinfo {author} {\bibfnamefont {J.~R.~V.}\ \bibnamefont {de~Aldana}},
  \bibinfo {author} {\bibfnamefont {L.}~\bibnamefont {Plaja}}, \bibinfo
  {author} {\bibfnamefont {G.~F.}\ \bibnamefont {Calvo}}, \ and\ \bibinfo
  {author} {\bibfnamefont {L.}~\bibnamefont {Roso}},\ }\href {\doibase
  10.1364/OE.18.003660} {\bibfield  {journal} {\bibinfo  {journal} {Opt.
  Express}\ }\textbf {\bibinfo {volume} {18}},\ \bibinfo {pages} {3660}
  (\bibinfo {year} {2010}{\natexlab{b}})}\BibitemShut {NoStop}%
\bibitem [{\citenamefont {Bond}\ \emph {et~al.}(2011)\citenamefont {Bond},
  \citenamefont {Fulda}, \citenamefont {Carbone}, \citenamefont {Kokeyama},\
  and\ \citenamefont {Freise}}]{LG_expression}%
  \BibitemOpen
  \bibfield  {author} {\bibinfo {author} {\bibfnamefont {C.}~\bibnamefont
  {Bond}}, \bibinfo {author} {\bibfnamefont {P.}~\bibnamefont {Fulda}},
  \bibinfo {author} {\bibfnamefont {L.}~\bibnamefont {Carbone}}, \bibinfo
  {author} {\bibfnamefont {K.}~\bibnamefont {Kokeyama}}, \ and\ \bibinfo
  {author} {\bibfnamefont {A.}~\bibnamefont {Freise}},\ }\href {\doibase
  10.1103/PhysRevD.84.102002} {\bibfield  {journal} {\bibinfo  {journal} {Phys.
  Rev. D}\ }\textbf {\bibinfo {volume} {84}},\ \bibinfo {pages} {102002}
  (\bibinfo {year} {2011})}\BibitemShut {NoStop}%
\bibitem [{\citenamefont {Klimov}\ \emph {et~al.}(2012)\citenamefont {Klimov},
  \citenamefont {Bloch}, \citenamefont {Ducloy},\ and\ \citenamefont
  {Rios~Leite}}]{LG_expression2}%
  \BibitemOpen
  \bibfield  {author} {\bibinfo {author} {\bibfnamefont {V.~V.}\ \bibnamefont
  {Klimov}}, \bibinfo {author} {\bibfnamefont {D.}~\bibnamefont {Bloch}},
  \bibinfo {author} {\bibfnamefont {M.}~\bibnamefont {Ducloy}}, \ and\ \bibinfo
  {author} {\bibfnamefont {J.~R.}\ \bibnamefont {Rios~Leite}},\ }\href
  {\doibase 10.1103/PhysRevA.85.053834} {\bibfield  {journal} {\bibinfo
  {journal} {Phys. Rev. A}\ }\textbf {\bibinfo {volume} {85}},\ \bibinfo
  {pages} {053834} (\bibinfo {year} {2012})}\BibitemShut {NoStop}%
\bibitem [{\citenamefont {Schiff}(1949)}]{schiff}%
  \BibitemOpen
  \bibfield  {author} {\bibinfo {author} {\bibfnamefont {L.}~\bibnamefont
  {Schiff}},\ }\href@noop {} {\emph {\bibinfo {title} {Quantum Mechanics}}}\
  (\bibinfo  {publisher} {McGraw-Hill Book Company, New York},\ \bibinfo {year}
  {1949})\BibitemShut {NoStop}%
\bibitem [{\citenamefont {Deiglmayr}\ \emph {et~al.}(2006)\citenamefont
  {Deiglmayr}, \citenamefont {Reetz-Lamour}, \citenamefont {Amthor},
  \citenamefont {Westermann}, \citenamefont {de~Oliveira},\ and\ \citenamefont
  {Weidemuller}}]{scaling}%
  \BibitemOpen
  \bibfield  {author} {\bibinfo {author} {\bibfnamefont {J.}~\bibnamefont
  {Deiglmayr}}, \bibinfo {author} {\bibfnamefont {M.}~\bibnamefont
  {Reetz-Lamour}}, \bibinfo {author} {\bibfnamefont {T.}~\bibnamefont
  {Amthor}}, \bibinfo {author} {\bibfnamefont {S.}~\bibnamefont {Westermann}},
  \bibinfo {author} {\bibfnamefont {A.}~\bibnamefont {de~Oliveira}}, \ and\
  \bibinfo {author} {\bibfnamefont {M.}~\bibnamefont {Weidemuller}},\ }\href
  {\doibase http://dx.doi.org/10.1016/j.optcom.2006.02.058} {\bibfield
  {journal} {\bibinfo  {journal} {Optics Communications}\ }\textbf {\bibinfo
  {volume} {264}},\ \bibinfo {pages} {293 } (\bibinfo {year}
  {2006})}\BibitemShut {NoStop}%
\bibitem [{\citenamefont {Wigner}(1959)}]{wigner3j}%
  \BibitemOpen
  \bibfield  {author} {\bibinfo {author} {\bibfnamefont {E.}~\bibnamefont
  {Wigner}},\ }\href@noop {} {\emph {\bibinfo {title} {Group Theory and Its
  Application to the Quantum Mechanics of Atomic Spectra}}}\ (\bibinfo
  {publisher} {New York: Academic Press},\ \bibinfo {year} {1959})\BibitemShut
  {NoStop}%
\bibitem [{\citenamefont {Brink}\ and\ \citenamefont {Satchler}(1962)}]{brink}%
  \BibitemOpen
  \bibfield  {author} {\bibinfo {author} {\bibfnamefont {D.}~\bibnamefont
  {Brink}}\ and\ \bibinfo {author} {\bibfnamefont {G.}~\bibnamefont
  {Satchler}},\ }\href@noop {} {\emph {\bibinfo {title} {Angular Momentum}}}\
  (\bibinfo  {publisher} {Oxford},\ \bibinfo {year} {1962})\BibitemShut
  {NoStop}%
\bibitem [{\citenamefont {Edmonds}(1968)}]{edmonds}%
  \BibitemOpen
  \bibfield  {author} {\bibinfo {author} {\bibfnamefont {A.}~\bibnamefont
  {Edmonds}},\ }\href@noop {} {\emph {\bibinfo {title} {Angular Momentum in
  Quantum Mechanics}}}\ (\bibinfo  {publisher} {Princeton University Press},\
  \bibinfo {year} {1968})\BibitemShut {NoStop}%
\end{thebibliography}%
\end{document}